\documentclass[aps,superscriptaddress,twocolumn,showpacs,dvips]{revtex4}
\usepackage{feynmp}
\usepackage{amssymb}
\usepackage{epsfig}
\usepackage{graphicx}
\usepackage{subfigure}
\usepackage{mathrsfs}
\usepackage{hyperref}
\usepackage{cases}

\begin{document}

\title{Renormalization group analysis of competition between distinct order parameters}

\author{Jing Wang}
\affiliation{Department of Modern Physics, University of Science and
Technology of China, Hefei, Anhui 230026, P.R. China}
\affiliation{National Synchrotron Radiation Laboratory, University
of Science and Technology of China, Hefei, Anhui 230029, P.R. China}
\author{Guo-Zhu Liu}
\affiliation{Department of Modern Physics, University of Science and
Technology of China, Hefei, Anhui 230026, P.R. China}

\begin{abstract}
We perform a detailed renormalization group analysis to study a
(2+1)-dimensional quantum field theory that is composed of two
interacting scalar bosons, which represent the order parameters for
two continuous phase transitions. This sort of field theory can
describe the competition and coexistence between distinct long-range
orders, and therefore plays a vital role in statistical physics and
condensed matter physics. We first derive and solve the
renormalization group equations of all the relevant physical
parameters, and then show that the system does not have any stable
fixed point in the lowest energy limit. Interestingly, this
conclusion holds in both the ordered and disordered phases, and also
at the quantum critical point. Therefore, the originally continuous
transitions are unavoidably turned to first-order due to ordering
competition. Moreover, we examine the impacts of massless Goldstone
boson generated by continuous symmetry breaking on ordering
competition, and briefly discuss the physical implications of our
results.
\end{abstract}

\pacs{11.10.-z, 11.10.Hi, 05.70.Fh}

\maketitle


\section{Introduction}

According to Landau, any continuous (second order) phase transition
can be described by defining some order parameter $\phi$, which
vanishes in the disordered phase but develops a finite vacuum
expectation value, $\langle \phi \rangle \neq 0$, in the ordered
phase~\cite{Zinn-Justin2002Book}. The finite $\langle \phi \rangle$
spontaneously breaks either a continuous or a discrete symmetry, and
is known to be associated with some long-range order. Classical phase
transitions always occur at certain critical temperature due to thermal
fluctuation, whereas quantum phase transitions~\cite{Sachdev1999Book}
take place at absolutely zero temperature driven by quantum fluctuation
and tuned by some external parameter, such as pressure and magnetic
field. No matter classical or quantum, phase transitions and the
associated critical behaviors are governed by an effective quantum
field theory of order parameter $\phi$~\cite{Zinn-Justin2002Book,
Kleinert2001Book}.

\begin{figure}
   \centering
   \includegraphics[width=2.6in]{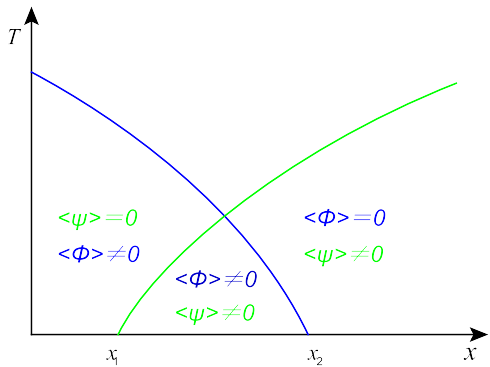}
   \vspace{-0.35cm}
\caption{Schematic phase diagram on $(x,T)$-plane of the order
parameters described by complex ($\psi$) and real ($\phi$) fields.
$x$ represents certain adjustable parameter whose variation tunes
phase transitions. $x_1$ and $x_2$ are quantum critical points of
the corresponding phase transitions,
respectively.}\label{Fig_coexist_comp_real}
\end{figure}

More interesting physics emerges when two or more long-range orders
coexist in one system~\cite{Sachdev2000Science}. This phenomenon is
indeed realized in a number of condensed matter systems, and thus
deserves careful and systematic investigations from viewpoints of
both statistical physics and quantum field theory. For instance,
high-$T_c$ cuprate superconductors may exhibit antiferromagnetic,
superconducting, and nematic long-range orders, depending on the
values of several tuning parameters~\cite{Kivelson03,Vojta}. These
orders are not independent. Instead, they compete strongly with each
other and under certain conditions can coexist homogeneously, giving
rise to rich properties. To illustrate the interplay between
distinct orders, we plot in Fig.~\ref{Fig_coexist_comp_real} a
schematic phase diagram defined on the $T-x$ plane, where $T$
denotes temperature and $x$ a free parameter that tunes phase
transitions. Here, $x_1$ and $x_2$ represent the quantum critical
points for two competing orders. These two orders coexist at zero
temperature in the region of $x_1 < x < x_2$. In terms of quantum
field theory, the ordering competition can be described by
constructing an effective model that is composed of two (or even
more) interacting scalar bosons~\cite{Arovas, Demler, She, Nussinov,
Millis10, Chowdhury, Schmalian}.

This sort of field theory is interesting for two reasons. First, it can be
applied to study the interplay between distinct orders and its physical
consequences in a number of realistic condensed matter systems, including
high temperature superconductors~\cite{Arovas,Demler,Metzner,Wang2014PRB,
Liu2012PRB,Wang2013NJP}, iron-based superconductors~\cite{Schmalian,
Fernandes2013PRL,Fernandes-Mills2013PRL,Chowdhury2013PRL}, and spinor
Bose-Einstein condensate~\cite{Stamper-Kurn_Ueda}. Second, within this
field theory, it was found that the strong interaction between two scalar
bosons can result in nontrivial properties, such as the general tendency
towards first-order transition~\cite{She,Millis10} and the occurrence of
nonuniform glassy phases~\cite{Nussinov}.

The symmetries that are spontaneously broken in various physical
problems usually fall into three categories: discrete symmetry,
continuous global symmetry, and continuous local symmetry. For
example, the transition from a uniform liquid to a nematic state is
known to be of Ising-type and the $C_4$ symmetry is broken down to
$C_2$ symmetry~\cite{Kivelson03,Vojta}. The corresponding order
parameter is a real scalar field. Formation of ferromagnetism and
antiferromagnetism break continuous rotational symmetries, and as
such generate massless Goldstone bosons (spin waves). In addition,
Bose-Einstein condensation of neutral bosons spontaneously break a
global U(1) symmetry, which also leads to massless Goldstone boson
(phonon). In BCS theory of superconductors, the formation of Cooper
pairs dynamically breaks the local U(1) gauge symmetry. However,
there is indeed no Goldstone boson in this case because it is
absorbed by the gauge field coupled to charged Cooper pairs. As a
result, the originally massless gauge boson becomes massive, which
is nothing but the Anderson-Higgs mechanism. In some peculiar
systems, ferromagnetism can compete and coexist with Bose-Einstein
condensate~\cite{Stamper-Kurn_Ueda, GuBongsBongs}, or with
superconductivity~\cite{BECSC_1, BECSC_2}. It is also possible that
superconductivity competes and coexists with nematic or
antiferromagnetic order~\cite{Kivelson03, Vojta}.

A natural question is how the ordering competition is influenced by
various symmetry-breaking patterns. Moreover, the order parameter
exhibits different properties in the disordered phase, ordered
phase, and quantum critical region~\cite{Vojta2003RPP,Sachdev2011PT}
(the small region on phase diagram around quantum critical point).
In the disordered phase, the order parameter has vanishing mean value,
and its quantum fluctuation is not expected to be strong. In the close
vicinity of quantum critical point, the mean value of order parameter
still vanishes, but the quantum fluctuation becomes singular and can
cause nontrivial quantum critical phenomena~\cite{Sachdev1999Book}.
In the ordered phase, the properties of order parameter is heavily
affected by the nature of broken symmetry. In the special case of
continuous symmetry breaking, the amplitude fluctuation of order
parameter is gapped (massive), whereas Goldstone bosonic excitation
is always gapless (massless). It is thus necessary to examine whether
Goldstone bosons play crucial roles in the description of ordering
competition. These problems were not systematically addressed
previously, which motivated us to revisit the problem of ordering
competition.

In this paper, we will study these problems within an effective
quantum field theory for the competition between two distinct order
parameters. As aforementioned, the order competition problem is
complicated and determined by the concrete symmetry-breaking
pattern. Moreover, the properties may be very different in ordered
and disordered phases. It is hardly possible to make a general
field-theoretic analysis that applies to all cases. For
concreteness, we only consider a particular case in which the system
contains one complex scalar field and one real scalar field. Our
focus will be on the stability of the system in the low-energy
region, which is usually examined by determining the possible stable
fixed points due to interactions. It would be easy to apply the same
scheme to study other systems of ordering competition.

The most suitable method to address the above issues is to perform
renormalization group (RG) calculations \cite{Wilson1975RMP}. We will
adopt the momentum-shell RG scheme~\cite{Polchinski,Shankar}, which
is physically intuitive and also formally simple. We first derive
the RG flow equations for all the relevant parameters in the field
theory, and then solve these self-consistently coupled equations
numerically. Interestingly, we find the interacting system does not
have any stable fixed point, which implies that the continuous phase
transition are turned to first-order. Such instability is primarily
driven by the quantum fluctuation of the amplitude of order parameter,
rather than Goldstone boson, and the competitive interaction between
distinct long-range orders.

The rest of the paper is organized as follows. We present the
quantum field theory and the corresponding Feynman rules in Sec.
\ref{Sec_eff_theory}. RG calculations are carried out in Sec.
\ref{Sec_RG_Calculations}. We briefly discuss the physical
implications of our results in Sec. \ref{Sec_RG_Solutions}. To
better understand the consequence of ordering competition, we
consider the region where two competing orders coexist homogeneously
in Sec. \ref{Sec_Coexisting_region}, and find that the originally
massless Goldstone boson become massive as a direct and nontrivial
consequence of ordering competition. In this case, the system also
undergoes a first-order instability. In Sec. \ref{Sec_Summary}, we
briefly summarize the results and discuss the possible extension
of the work.

\section{Effective field theory for ordering competition}\label{Sec_eff_theory}

In principle, an order parameter can be a real scalar field, a
complex scalar field, or a vector field. In this paper, we are
mainly interested in the interplay between two distinct scalar
fields. To keep a balance between generality and simplicity, we
assume one of them is a complex scalar field whereas the other a
real scalar field. Moreover, we consider a (2+1)-dimensional model
since the ordering competition phenomena usually take place in
layered superconductors. It is straightforward to generalize the
analysis to other forms of order parameters and to other space-time
dimensions.

The competition between two distinct order parameters can be
described by the following field theory
\begin{eqnarray}
\mathcal{L}&=&\mathcal{L}_\psi + \mathcal{L}_\phi
+\mathcal{L}_{\psi\phi},\label{Eq_L_total}\\
\mathcal{L}_\psi&=&\partial_\mu \psi^\dagger\partial_\mu \psi
-\alpha |\psi|^2+\frac{\beta}{2}|\psi|^4,\label{Eq_L_psi}\\
\mathcal{L}_\phi&=&\frac{1}{2}(\partial_\mu \phi)^2
+r\phi^2 +\frac{u}{2}\phi^4,\label{Eq_L_phi}\\
\mathcal{L}_{\psi\phi}&=&\lambda|\psi|^2\phi^2,\label{Eq_L_psi_phi}
\end{eqnarray}
where $\mathcal{L}_\psi$ and $\mathcal{L}_\phi$ are the
Ginzburg-Landau model for order parameters $\psi$ and $\phi$,
respectively. $\mathcal{L}_{\psi\phi}$ represents the interaction
between $\psi$ and $\phi$. Such interaction is repulsive or
competitive if the coupling constant $\lambda$ is chosen to be
positive. The mass parameters $\alpha$ and $r$ tune the phase
transitions that lead to finite mean values of $\psi$ and $\phi$,
respectively. For example, if $\alpha < 0$, the free energy of
$\psi$ exhibits its minimum at $\langle \psi\rangle = 0$, which
implies the system is in the disordered phase. If $\alpha
> 0$, the minimum of free energy of $\psi$ is located at a finite
$\langle \psi \rangle$, so the system is in the ordered phase. It is
therefore clear that $\alpha = 0$ represents the zero temperature
quantum critical point that separates disordered and ordered phases,
corresponding to $x_1$ in the phase diagram Fig.~\ref{Fig_coexist_comp_real}.
Analogously, $r=0$ is the quantum critical point for order parameter $\phi$,
represented by $x_2$ in Fig.~\ref{Fig_coexist_comp_real}. In this paper, we
assume that $\alpha > 0$ and $r > 0$. In addition, $\beta$ and $u$ are both
positive according to the standard theory of continuous phase transition.

Let us assume that $\psi$ is a complex scalar field, which may be
the order parameter of superfluidity, superconductivity, or
antiferromagnetism. In the vicinity of quantum critical point $x_2$,
the field $\psi$ acquires a finite vacuum expectation value due to
vacuum degeneracy, i.e.,
\begin{eqnarray}
\langle\psi\rangle = V_0 = \sqrt{\frac{\alpha}{\beta}}.
\end{eqnarray}
Quantum fluctuation of $\psi$ is known to be strong at zero
temperature, especially in the nearby of quantum critical point. The
fluctuation of $\psi$ around its mean value can be described by
introducing two new fields $h$ and $\eta$~\cite{Kleinert2003NPB},
\begin{eqnarray}
\psi = V_0+\frac{1}{\sqrt{2}}(h+i\eta),\label{Eq_psi_renormalized}
\end{eqnarray}
where
\begin{eqnarray}
\langle h \rangle=\langle \eta \rangle=0.
\end{eqnarray}
In previous analysis of ordering competition, the influence of quantum fluctuation of
order parameter in the ordered phase is not carefully analyzed, and it is unclear whether
massless Goldstone boson plays an important role. By employing the field parametrization Eq.~(\ref{Eq_psi_renormalized}), we are allowed to separate the contributions of amplitude
fluctuation of order parameter and Goldstone boson. In order to make the impact of Goldstone
boson more transparent, we assume $\phi$ is a real scalar field that is induced by discrete
symmetry breaking and therefore does not contain Goldstone boson.

\begin{figure}
   \centering
   \includegraphics[width=2.85in]{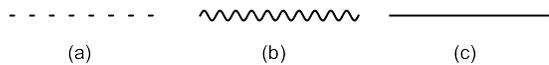}
   \vspace{-0.3cm}
\caption{Free propagators for (a) $h$, (b) $\eta$, and (c)
$\phi$.}\label{Fig_propagators_comp_real}
\end{figure}

\begin{figure}
   \centering
   \includegraphics[width=2.6in]{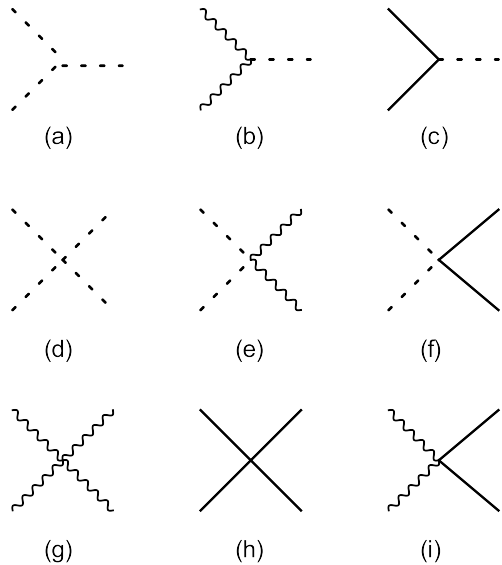}
   \vspace{-0.3cm}
\caption{Vertices: (a) $h^3$; (b) $\eta^2h$; (c) $\phi^2 h$; (d) $h^4$;
(e) $h^2\eta^2$; (f) $h^2\phi^2$; (g) $\eta^4$; (h) $\phi^4$;
(i) $\eta^2\phi^2$.}\label{Fig_vertex_comp_real}
\end{figure}

Substituting Eq.~(\ref{Eq_psi_renormalized}) into Eq.~(\ref{Eq_L_total}),
we are left with the following Lagrangian density
\begin{eqnarray}
\mathcal{L}_{\mathrm{eff}} &=& \frac{1}{2}(\partial_\mu
h)^2+\alpha_h h^2+\gamma_hh^3 +\frac{\beta_h}{2}h^4 +
\frac{1}{2}(\partial_\mu\eta)^2 \nonumber\\
&& +\frac{\beta_\eta}{2}\eta^4 + \frac{1}{2}(\partial_\mu\phi)^2
+\alpha_\phi\phi^2 +\frac{\beta_\phi}{2}\phi^4 + \gamma_{\eta
h}\eta^2h \nonumber\\
&&+\gamma_{\phi h}\phi^2h+\lambda_{\eta h}h^2\eta^2 +
\lambda_{h\phi}h^2\phi^2 +\lambda_{\eta\phi}\eta^2\phi^2,
\label{Eq_effective_L}
\end{eqnarray}
where the new parameters are defined as
\begin{eqnarray}
\alpha_h &=& \alpha,\,\,\,\,\,\,\alpha_\phi =
\frac{\lambda\alpha}{\beta}+r, \label{Eq_para_trans1}\\
\beta_h &=& \frac{\beta}{4},\,\,\,\beta_\eta =
\frac{\beta}{4},\,\,\,\beta_\phi = u, \label{Eq_para_trans2}\\
\gamma_h &=& \frac{\sqrt{2\alpha\beta}}{2},\gamma_{\eta h} =
\frac{\sqrt{2\alpha\beta}}{2},\gamma_{\phi h} = \lambda
\sqrt{\frac{2\alpha}{\beta}}, \label{Eq_para_trans3}\\
\lambda_{\eta h} &=& \frac{\beta}{4},\,\,\,\lambda_{h\phi} =
\frac{\lambda}{2}, \,\,\,\lambda_{\eta\phi} =
\frac{\lambda}{2}.\label{Eq_para_trans4}
\end{eqnarray}
From Eq.~(\ref{Eq_effective_L}), it is easy to extract the free
propagators of fields $h$, $\eta$, and $\phi$, namely
\begin{eqnarray}
G_h(k)&=&\frac{1}{k^2+2\alpha_h},\\
G_\eta(k)&=&\frac{1}{k^2},\label{Eq_propagator_eta}\\
G_\phi(k)&=&\frac{1}{k^2+2\alpha_\phi}.
\end{eqnarray}
The corresponding Feynman rules for free propagators and free
vertices are shown in Fig.~\ref{Fig_propagators_comp_real} and
Fig.~\ref{Fig_vertex_comp_real}. The Goldstone boson generated by
continuous symmetry breaking is represented by $\eta$, whose
masslessness can be readily seen from both Eq.~(\ref{Eq_effective_L})
and Eq.~(\ref{Eq_propagator_eta}). On the other hand, the amplitude
fluctuation of order parameter $\psi$ is encoded in field $h$, which
is massive when $x \neq x_1$.

In many field-theoretic treatments of phase transition, especially
in the context of condensed matter systems, the amplitude
fluctuation of order parameter, $h$ in our case, is usually
considered as unimportant and hence omitted. At the mean-field
level, this approximation is expected to perfectly valid. However,
at zero temperature, the quantum fluctuation of $h$ is important and
has unnegligible effect, which makes the mean-field treatment
unreliable. This effect becomes more and more significant as one
approaches the quantum critical point, where the quantum fluctuation
of order parameter is indeed singular. As will be shown below, the
amplitude fluctuation $h$ and its coupling with competing order is
able to drive an instability of the system. In Ref.~\cite{Kleinert2003NPB},
Kleinert and Nogueira investigated the interaction between a superconducting
order parameter and an abelian gauge field, where the field parametrization Eq.~(\ref{Eq_psi_renormalized}) of complex order parameter was adopted, and
obtained an infrared-stable fixed point by means of RG method. It is also
interesting to notice that, a recent work \cite{Benfatto} studied the impact
of amplitude fluctuation (Higgs mode) in a system with coexisting
superconducting and charge-density-wave orders, and revealed
important observable effects of amplitude fluctuation.

\section{Renormalization group analysis}\label{Sec_RG_Calculations}

There are several RG schemes available in the literature, ranging from
Wilson's original momentum-shell scheme~\cite{Wilson1975RMP} to the more
complicated function RG scheme~\cite{Wetterich,Metzner2012RMP}. Here we
adopt the momentum-shell scheme~\cite{Polchinski,Shankar}.

\subsection{Effective action}

The essence of RG analysis is to integrate out high energy (small
scale) degrees of freedom and select out low energy (large scale)
degrees of freedom. It is therefore necessary to express the action
of field operators $h$, $\eta$, and $\phi$ as integrals over momenta
and energies. Formally, we have
\begin{eqnarray}
S&=&S_{h}+S_{\eta}+S_{\phi}+S_{\eta^2 h}+S_{\phi^2h}+
S_{h^2\eta^2}\nonumber\\
&&+S_{h^2\phi^2}+S_{\eta^2\phi^2},\label{Eq_action_eff}
\end{eqnarray}
where
\begin{eqnarray}
S_{h}&=&\frac{1}{2}\int \frac{d^2\mathbf{k}d\omega}{(2\pi)^3}
\left(2\alpha_h+\mathbf{k}^2+\omega^2\right)h^2\nonumber\\
&&+\gamma_h\int\prod^3_{m=1}\frac{d^2\mathbf{k}_md\omega_m}
{(2\pi)^3}\Delta(m)h^3 \nonumber\\
&&+\frac{\beta_h}{2}\int\prod^4_{m=1}\frac{d^2\mathbf{k}_md\omega_m}
{(2\pi)^3}\Delta(m)h^4,\label{Eq_action_eff_h}\\
S_{\eta}&=&\frac{1}{2}\int \frac{d^2\mathbf{k}d\omega}{(2\pi)^3}
\left(\mathbf{k}^2+\omega^2\right)\eta^2 \nonumber\\
&&+\frac{\beta_\eta}{2}\int\prod^4_{m=1}\frac{d^2\mathbf{k}_md\omega_m}
{(2\pi)^3}\Delta(m)\eta^4,\\
S_{\phi}&=&\int \frac{d^2\mathbf{q}d\epsilon}{(2\pi)^3}
\frac{1}{2}\left(2\alpha_\phi+\mathbf{q}^2+\epsilon^2\right)\phi^2 \nonumber\\
&&+\frac{\beta_\phi}{2} \int \prod^4_{m=1}\frac{d^2\mathbf{q}_md\epsilon_m}
{(2\pi)^3}\Delta(m)\phi^4,
\end{eqnarray}
and
\begin{eqnarray}
S_{\eta^2 h}&=&\gamma_{\eta h}\int \prod_{i=1,2}
\frac{d^2\mathbf{k}_id\omega_i}{(2\pi)^3}
\frac{d^2\mathbf{k'}d\omega'}{(2\pi)^3} \nonumber\\
&&\times\eta(\mathbf{k}_i,\omega_i)h(\mathbf{k'},\omega')
\Gamma^{(\mathbf{k},\mathbf{k'})}_{(\omega,\omega')},\\
S_{\phi^2h}&=&\gamma_{\phi h}\int \prod_{i=1,2}
\frac{d^2\mathbf{q}_id\epsilon_i}{(2\pi)^3}
\frac{d^2\mathbf{k}d\omega}{(2\pi)^3} \nonumber\\
&&\times\phi(\mathbf{q}_i,\epsilon_i)h(\mathbf{k},\omega)
\Gamma^{(\mathbf{q},\mathbf{k})}_{(\epsilon,\omega)},\\
S_{h^2\eta^2}&=&\lambda_{\eta h}\int \prod_{i=1,2}
\frac{d^2\mathbf{k}_id\omega_id^2\mathbf{k'}_id\omega'_i}{(2\pi)^6}
\nonumber\\
&&\times\eta(\mathbf{k}_i,\omega_i)h(\mathbf{k'}_i,\omega'_i)
\Xi^{(\mathbf{k},\mathbf{k'})}_{(\omega,\omega')},\\
S_{h^2\phi^2}&=&\lambda_{h\phi}\int \prod_{i=1,2}
\frac{d^2\mathbf{k}_id\omega_id^2\mathbf{q}_id\epsilon_i}{(2\pi)^6}
\nonumber\\
&&\times h(\mathbf{k}_i,\omega_i)\phi(\mathbf{q}_i,\epsilon_i)
\Xi^{(\mathbf{k},\mathbf{q})}_{(\omega,\epsilon)},\\
S_{\eta^2\phi^2}&=&\lambda_{\eta\phi}\int \prod_{i=1,2}
\frac{d^2\mathbf{k}_id\omega_id^2\mathbf{q}_id\epsilon_i}{(2\pi)^6}
\nonumber\\
&&\times\eta(\mathbf{k}_i,\omega_i)\phi(\mathbf{q}_i,\epsilon_i)
\Xi^{(\mathbf{k},\mathbf{q})}_{(\omega,\epsilon)}).
\end{eqnarray}
Here, in order to simplify notations, we have defined
\begin{eqnarray}
\Delta(m) &\equiv& \delta^2\left(\sum\mathbf{k}_m\right)
\delta\left(\sum\omega_m\right), \nonumber \\
\Gamma^{(\mathbf{x},\mathbf{y})}_{(s,t)}&\equiv&\delta^2\left(\mathbf{x}_1
+\mathbf{x}_2+\mathbf{y}\right) \delta\left(s_1+s_2+t\right),\nonumber \\
\Xi^{(\mathbf{x},\mathbf{y})}_{(s,t)}&\equiv&\delta^2
\left(\mathbf{x}_1+\mathbf{x}_2
+\mathbf{y}_1+\mathbf{y}_2\right)\delta\left(s_1+s_2 + t_1 +
t_2\right).\nonumber
\end{eqnarray}
Since all the terms in $S$ are already written as integrals over
momenta and energies, we are now ready to eliminate the modes of
large momenta and high energies.

\subsection{Scaling transformations}

Following the formalism presented in Refs.~\cite{Polchinski,Shankar},
we first make scaling transformations,
\begin{eqnarray}
k_{x,y} &=& k'_{x,y}e^{-l},\\
\omega &=& \omega' e^{-l},\\
q_{x,y} &=& q'_{x,y}e^{-l},\\
\epsilon &=& \epsilon' e^{-l},
\end{eqnarray}
where $l$ is a running scale that goes to infinity at
the lowest energy. Under these transformations, the field operators
$h$, $\eta$, and $\phi$ should transform accordingly so that the
free parts of actions $S_{h}$, $S_{\eta}$, and $S_{\phi}$ remain
unchanged. In order words, they are defined as free fixed points
under RG scaling transformations. It is easy to know from the free
actions that $h$, $\eta$, and $\phi$ should be re-scaled as
\begin{eqnarray}
h(\mathbf{k},\omega) &=& h'(\mathbf{k'},\omega')e^{5l/2},\\
\eta(\mathbf{k},\omega) &=& \eta'(\mathbf{k'},\omega')e^{5l/2},\\
\phi(\mathbf{q},\epsilon) &=& \phi'(\mathbf{q'},\epsilon')e^{5l/2}.
\end{eqnarray}

\subsection{Slow and fast modes of field operators}

To proceed, we need to separate each field operator into slow mode
and fast mode, i.e.,
\begin{eqnarray}
h &=& h_s + h_f, \\
\eta &=& \eta_s + \eta_f, \\
\phi &=& \phi_s + \phi_f.
\end{eqnarray}
Such separation would be meaningless without specifying which modes
are fast or slow. We introduce an ultraviolet cutoff $\Lambda$,
which naturally exists in realistic condensed matter systems, and
then rescale momenta and energy using $\Lambda$, i.e.,
\begin{eqnarray}
\mathbf{k} \rightarrow \mathbf{k}/\Lambda,\hspace{0.25cm} \omega
\rightarrow \omega/\Lambda.
\end{eqnarray}
In terms of new variables, we define the slow modes as
\begin{eqnarray}
h_s &=& h(k)\hspace{0.25cm} \mathrm{for}\hspace{0.25cm}
0 < k < b,\\
\eta_s &=& \eta(k)\hspace{0.25cm} \mathrm{for}\hspace{0.25cm}
0 < k < b,\\
\phi_s &=& \phi(k)\hspace{0.25cm} \mathrm{for}\hspace{0.25cm} 0 < k
< b,
\end{eqnarray}
and the fast modes as
\begin{eqnarray}
h_f &=& h(k)\hspace{0.25cm} \mathrm{for} \hspace{0.25cm}
b < k < 1,\\
\eta_f &=& \eta(k)\hspace{0.25cm} \mathrm{for} \hspace{0.25cm}
b < k < 1,\\
\phi_f &=& \phi(k)\hspace{0.25cm} \mathrm{for} \hspace{0.25cm} b < k
< 1,
\end{eqnarray}
where
\begin{eqnarray}
b = e^{-l}.
\end{eqnarray}

\begin{figure}
   \centering
   \includegraphics[width=3.0in]{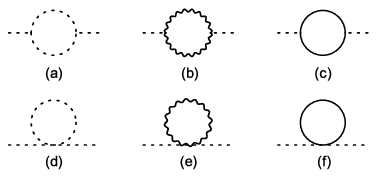}
   \vspace{-0.3cm}
\caption{One-loop corrections to $\alpha_h$.}\label{Fig_alpha_h}
\end{figure}
\begin{figure}
   \centering
   \includegraphics[width=1.9in]{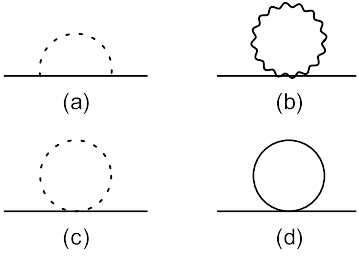}
   \vspace{-0.3cm}
\caption{One-loop corrections to
$\alpha_\phi$.}\label{Fig_alpha_phi}
\end{figure}

Based on the above mode separation, the whole action
Eq.~(\ref{Eq_action_eff}) can be decomposed into three parts:
$S^{s}$ that contains only slow modes, $S^{f}$ that contains only
fast modes, and $S^{sf}$ that contains both slow and fast modes. The
action can be rewritten in the following form
\begin{eqnarray}
S&=&S^{s}+S^{f}+S^{sf}, \label{Eq_action_eff_s_f} \\
S^{s}&=&S^s_{h}+S^s_{\eta}+S^s_{\phi}+S^s_{\eta^2 h}+S^s_{\phi^2h}\nonumber\\
&&+S^s_{h^2\eta^2}+S^s_{h^2\phi^2}+S^s_{\eta^2\phi^2}, \\
S^{f}&=&S^f_{h}+S^f_{\eta}+S^f_{\phi}+S^f_{\eta^2 h}+S^f_{\phi^2h}\nonumber\\
&&+S^f_{h^2\eta^2}+S^f_{h^2\phi^2}+S^f_{\eta^2\phi^2}, \\
S^{sf}&=&S^{sf}_{h}+S^{sf}_{\eta}+S^{sf}_{\phi}+S^{sf}_{\eta
h}+S^{sf}_{h\phi}+S^{sf}_{\eta\phi}.
\end{eqnarray}
More concretely, the slow/fast parts are
\begin{eqnarray}
S^{s,f}_{h}&=&\frac{1}{2}\int \frac{d^2\mathbf{k}d\omega}{(2\pi)^3}
\left(2\alpha_h+\mathbf{k}^2+\omega^2\right)h^2_{s,f}\nonumber\\
&&+\gamma_h\int\prod^3_{m=1}\frac{d^2\mathbf{k}_md\omega_m}
{(2\pi)^3}\Delta(m)h^3_{s,f}\nonumber\\
&&+\frac{\beta_h}{2}\int\prod^4_{m=1}\frac{d^2\mathbf{k}_md\omega_m}
{(2\pi)^3}\Delta(m)h^4_{s,f},\\
S^{s,f}_{\eta}&=&\frac{1}{2}\int \frac{d^2\mathbf{k}d\omega}{(2\pi)^3}
\left(\mathbf{k}^2+\omega^2\right)\eta^2_{s,f}\nonumber\\
&&+\frac{\beta_\eta}{2}\int\prod^4_{m=1}\frac{d^2\mathbf{k}_md\omega_m}
{(2\pi)^3}\Delta(m)\eta^4_{s,f},\\
S^{s,f}_{\phi}&=&\frac{1}{2}\int \frac{d^2\mathbf{q}d\epsilon}{(2\pi)^3}
\left(2\alpha_\phi+\mathbf{q}^2+\epsilon^2\right)\phi^2_{s,f}\nonumber\\
&&+\frac{\beta_\phi}{2} \int \prod^4_{m=1}\frac{d^2\mathbf{q}_md\epsilon_m}
{(2\pi)^3}\Delta(m)\phi^4_{s,f},\\
S^{s,f}_{\eta^2 h}&=&\gamma_{\eta h}\int \prod_{i=1,2}
\frac{d^2\mathbf{k}_id\omega_i}{(2\pi)^3}
\frac{d^2\mathbf{k'}d\omega'}{(2\pi)^3}\nonumber\\
&&\times\eta_{s,f}(\mathbf{k}_i,\omega_i)h_{s,f}(\mathbf{k'},\omega')
\Gamma^{(\mathbf{k},\mathbf{k'})}_{(\omega,\omega')},\\
S^{s,f}_{\phi^2h}&=&\gamma_{\phi h}\int \prod_{i=1,2}
\frac{d^2\mathbf{q}_id\epsilon_i}{(2\pi)^3}
\frac{d^2\mathbf{k}d\omega}{(2\pi)^3}\nonumber\\
&&\times\phi_{s,f}(\mathbf{q}_i,\epsilon_i)h_{s,f}(\mathbf{k},\omega)
\Gamma^{(\mathbf{q},\mathbf{k})}_{(\epsilon,\omega)},\\
S^{s,f}_{h^2\eta^2}&=&\lambda_{\eta h}\int \prod_{i=1,2}
\frac{d^2\mathbf{k}_id\omega_i}{(2\pi)^3}\frac{d^2\mathbf{k'}_id\omega'_i}
{(2\pi)^3}\nonumber\\
&&\times\eta_{s,f}(\mathbf{k}_i,\omega_i)h_{s,f}(\mathbf{k'}_i,\omega'_i)
\Xi^{(\mathbf{k},\mathbf{k'})}_{(\omega,\omega')},\\
S^{s,f}_{h^2\phi^2}&=&\lambda_{h\phi}\int \prod_{i=1,2}
\frac{d^2\mathbf{k}_id\omega_i}{(2\pi)^3}\frac{d^2\mathbf{q}_id\epsilon_i}
{(2\pi)^3}\nonumber\\
&&\times h_{s,f}(\mathbf{k}_i,\omega_i)\phi_{s,f}(\mathbf{q}_i,\epsilon_i)
\Xi^{(\mathbf{k},\mathbf{q})}_{(\omega,\epsilon)},\\
S^{s,f}_{\eta^2\phi^2}&=&\lambda_{\eta\phi}\int \prod_{i=1,2}
\frac{d^2\mathbf{k}_id\omega_i}{(2\pi)^3}\frac{d^2\mathbf{q}_id\epsilon_i}
{(2\pi)^3}\nonumber\\
&&\times\eta_{s,f}(\mathbf{k}_i,\omega_i)\phi_{s,f}(\mathbf{q}_i,\epsilon_i)
\Xi^{(\mathbf{k},\mathbf{q})}_{(\omega,\epsilon)};
\end{eqnarray}
and the slow-fast mixing terms are
\begin{eqnarray}
S^{sf}_{h} &=&3\gamma_h\int\prod^3_{m=1}\frac{d^2\mathbf{k}_md\omega_m}
{(2\pi)^3}\Delta(m) \left(h_sh_fh_f+h_sh_sh_f\right) \nonumber\\
&&+3\beta_h\int\prod^4_{m=1}\frac{d^2\mathbf{k}_md\omega_m}
{(2\pi)^3}\Delta(m)h_sh_sh_fh_f,\\
S^{sf}_{\eta}
&=&3\beta_\eta\int\prod^4_{m=1}\frac{d^2\mathbf{k}_md\omega_m}
{(2\pi)^3}\Delta(m)\eta_s\eta_s\eta_f\eta_f,\\
S^{sf}_{\phi}&=&3\beta_\phi\int
\prod^4_{m=1}\frac{d^2\mathbf{q}_md\epsilon_m}
{(2\pi)^3}\Delta(m)\phi_s\phi_s\phi_f\phi_f,
\end{eqnarray}

\begin{widetext}
and
\begin{eqnarray}
S^{sf}_{\eta h}
&=&\int \prod_{i=1,2}
\frac{d^2\mathbf{k}_id\omega_i}{(2\pi)^3}
\frac{d^2\mathbf{k'}d\omega'}{(2\pi)^3}
\Bigl\{\gamma_{\eta h}\Bigl[\eta_s(\mathbf{k}_i,\omega_i)h_f(\mathbf{k'},\omega')
+\eta_f(\mathbf{k}_i,\omega_i)h_s(\mathbf{k'},\omega')\Bigr]\nonumber\\
&&+2\gamma_{\eta h}\eta_s(\mathbf{k}_1,\omega_1)\eta_f(\mathbf{k}_2,\omega_2)
h_f(\mathbf{k'},\omega')\Bigr\}\Gamma^{(\mathbf{k},\mathbf{k'})}_{(\omega,\omega')}
+\lambda_{\eta h}\int \prod_{i=1,2}\frac{d^2\mathbf{k}_id\omega_i}{(2\pi)^3}
\frac{d^2\mathbf{k'}_id\omega'_i}{(2\pi)^3}\nonumber\\
&&\times\Bigl[\eta_s(\mathbf{k}_i,\omega_i)h_f(\mathbf{k'}_i,\omega'_i)
+\eta_f(\mathbf{k}_i,\omega_i)h_s(\mathbf{k'}_i,\omega'_i)\Bigr]
\Xi^{(\mathbf{k},\mathbf{k'})}_{(\omega,\omega')}+4\lambda_{\eta
h}\int \prod_{i=1,2} \frac{d^2\mathbf{k}_id\omega_i}{(2\pi)^3}
\frac{d^2\mathbf{k'}_id\omega'_i}{(2\pi)^3}\nonumber\\
&&\times\eta_s(\mathbf{k}_1,\omega_1)
\eta_f(\mathbf{k}_2,\omega_2)h_s(\mathbf{k'}_1,\omega'_1)h_f(\mathbf{k'}_2,\omega'_2)
\Xi^{(\mathbf{k},\mathbf{k'})}_{(\omega,\omega')},\\
S^{sf}_{h\phi}
&=&\int \prod_{i=1,2}
\frac{d^2\mathbf{q}_id\epsilon_i}{(2\pi)^3}
\frac{d^2\mathbf{k}d\omega}{(2\pi)^3}
\Bigr\{\gamma_{\phi h}\Bigl[\phi_s(\mathbf{q}_i,\epsilon_i)h_f(\mathbf{k},\omega)
+\phi_f(\mathbf{q}_i,\epsilon_i)h_s(\mathbf{k},\omega)\Bigr]\nonumber\\
&&+2\gamma_{\phi h}\phi_s(\mathbf{q}_1,\epsilon_1)
\phi_f(\mathbf{q}_2,\epsilon_2)h_f(\mathbf{k},\omega)\Bigr\}
\Gamma^{(\mathbf{k},\mathbf{q})}_{(\epsilon,\omega)}
+\lambda_{h\phi}\int \prod_{i=1,2}\frac{d^2\mathbf{q}_id\epsilon_i}{(2\pi)^3}
\frac{d^2\mathbf{k}d\omega}{(2\pi)^3}\nonumber\\
&&\times\Bigl[h_s(\mathbf{k}_i,\omega_i)\phi_f(\mathbf{q}_i,\epsilon_i)
+h_f(\mathbf{k}_i,\omega_i)\phi_s(\mathbf{q}_i,\epsilon_i)\Bigr]
\Xi^{(\mathbf{k},\mathbf{q})}_{(\omega,\epsilon)}+4\lambda_{h\phi}\int
\prod_{i=1,2}\frac{d^2\mathbf{q}_id\epsilon_i}{(2\pi)^3}
\frac{d^2\mathbf{k}d\omega}{(2\pi)^3}\nonumber\\
&&\times h_s(\mathbf{k}_1,\omega_1)h_f(\mathbf{k}_2,\omega_2)\phi_s(\mathbf{q}_1,\epsilon_1)
\phi_f(\mathbf{q}_2,\epsilon_2)\Xi^{(\mathbf{k},\mathbf{q})}_{(\omega,\epsilon)},\\
S^{sf}_{\eta\phi}
&=&\lambda_{\eta\phi}\int \prod_{i=1,2}
\frac{d^2\mathbf{k}_id\omega_i}{(2\pi)^3}\frac{d^2\mathbf{q}_id\epsilon_i}
{(2\pi)^3}\Bigl[\eta_s(\mathbf{k}_i,\omega_i)\phi_f(\mathbf{q}_i,\epsilon_i)
+\eta_f(\mathbf{k}_i,\omega_i)\phi_s(\mathbf{q}_i,\epsilon_i)\Bigr]
\Xi^{(\mathbf{k},\mathbf{q})}_{(\omega,\epsilon)}\nonumber\\
&&+4\lambda_{\eta\phi}\int \prod_{i=1,2}\frac{d^2\mathbf{k}_id\omega_i}{(2\pi)^3}
\frac{d^2\mathbf{q}_id\epsilon_i}{(2\pi)^3}\eta_s(\mathbf{k}_1,\omega_1)
\eta_f(\mathbf{k}_2,\omega_2)\phi_s(\mathbf{q}_1,\epsilon_1)
\phi_f(\mathbf{q}_2,\epsilon_2)\Xi^{(\mathbf{k},\mathbf{q})}_{(\omega,\epsilon)}.
\end{eqnarray}
\end{widetext}

After carrying out the above mode decomposition, we can now write
the partition function as
\begin{eqnarray}
Z &=& \int \mathcal{D}h_s\mathcal{D}\eta_s \mathcal{D}\phi_s
\mathcal{D}h_f\mathcal{D}\eta_f \mathcal{D}\phi_f\nonumber\\
&&\times\exp\left({S^{s}+S^{f} +S^{sf}}\right).
\end{eqnarray}
We then integrate over $\eta^f$, $h^f$, and $\phi^f$, and have
\begin{eqnarray}
Z&=&\int \mathcal{D}h_s\mathcal{D}\eta_s \mathcal{D}\phi_s
\mathrm{exp}\left(S'^s_{h} + S'^s_{\eta} + S'^s_{\phi} +
S'^s_{\eta^2 h}\right. \nonumber\\
&&\left.+S'^s_{\phi^2h}+S'^s_{h^2\eta^2}+S'^s_{h^2\phi^2}
+S'^s_{\eta^2\phi^2}\right)\nonumber\\
&\equiv& \int \mathcal{D}h_s\mathcal{D}\eta_s
\mathcal{D}\phi_s\exp{\left(S'_{\mathrm{eff}}\right)}.
\end{eqnarray}
The next step is to integrate over all the fast modes, followed by
scaling transformations. Then an effective action will be obtained.
The functional integration can be performed by using the standard
perturbation expansion, with the help of the following
identity~\cite{Shankar}
\begin{eqnarray}
\exp\left(-S_{\mathrm{eff}}\right)=\exp\left(-\langle
S_c\rangle_f+\frac{1}{2}\langle S^2_c\rangle_f\right),
\end{eqnarray}
where $S^2_c$ corresponds to the connected average.  In the next
subsection, we will calculate RG equations up to one-loop level in
powers of small coupling parameters.

\subsection{One-loop corrections}

All the one-loop correction diagrams are plotted in
Figs.~\ref{Fig_alpha_h}~-~\ref{Fig_lambda_eta_phi}. As depicted
in Fig.~\ref{Fig_alpha_h}, there are six one-loop diagrams
contributing to the renormalization of $\alpha_h$. One should
calculate them one by one so as to get the one-loop correction to
$\alpha_h$. The contribution from diagrams presented in the first
line of Fig.~\ref{Fig_alpha_h} is
\begin{widetext}

\begin{figure}[htb]
\centering \hspace{-1.2cm}
\begin{minipage}{5.5cm}
\includegraphics[width=1.9in]{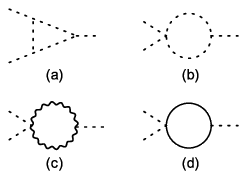}
\vspace{-0.3cm} \caption{One-loop corrections to
$\gamma_h$.}\label{Fig_gamma_h}
\end{minipage}
\hspace{0.1cm}
\begin{minipage}{5.8cm}
\includegraphics[width=1.9in]{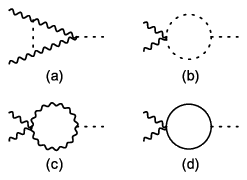}
\vspace{-0.3cm} \caption{One-loop corrections to $\gamma_{\eta
h}$.}\label{Fig_gamma_eta_h}
\end{minipage}
\hspace{0.1cm}
\begin{minipage}{5.8cm}
\includegraphics[width=1.9in]{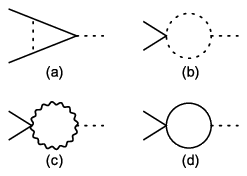}
\vspace{-0.3cm} \caption{One-loop corrections to $\gamma_{\phi
h}$.}\label{Fig_gamma_phi_h}
\end{minipage}
\end{figure}

\begin{figure}[htb]
\centering
\hspace{-1.2cm}
\begin{minipage}{5.5cm}
\includegraphics[width=1.9in]{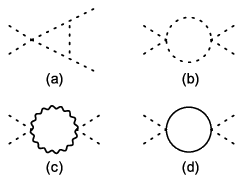}
\vspace{-0.3cm}
\caption{One-loop corrections to $\beta_h$.}\label{Fig_beta_h}
\end{minipage}
\hspace{0.1cm}
\begin{minipage}{5.8cm}
\includegraphics[width=1.9in]{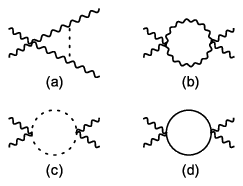}
\vspace{-0.3cm}
\caption{One-loop corrections to $\beta_\eta$.}\label{Fig_beta_eta}
\end{minipage}
\hspace{0.1cm}
\begin{minipage}{5.8cm}
\includegraphics[width=1.9in]{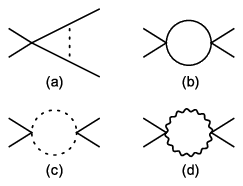}
\vspace{-0.3cm}
\caption{One-loop corrections to $\beta_\phi$.}\label{Fig_beta_phi}
\end{minipage}

\end{figure}
\begin{figure}[htb]
\centering
\hspace{-1.2cm}
\begin{minipage}{5.8cm}
\includegraphics[width=1.88in]{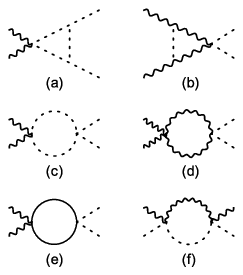}
\vspace{-0.3cm}
\caption{One-loop corrections to $\lambda_{\eta h}$.}\label{Fig_lambda_eta_h}
\end{minipage}
\hspace{0.1cm}
\begin{minipage}{5.8cm}
\includegraphics[width=1.88in]{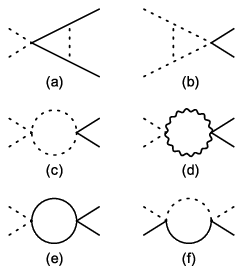}
\vspace{-0.3cm}
\caption{One-loop corrections to $\lambda_{h\phi}$.}\label{Fig_lambda_h_phi}
\end{minipage}
\hspace{0.1cm}
\begin{minipage}{5.8cm}
\includegraphics[width=1.88in]{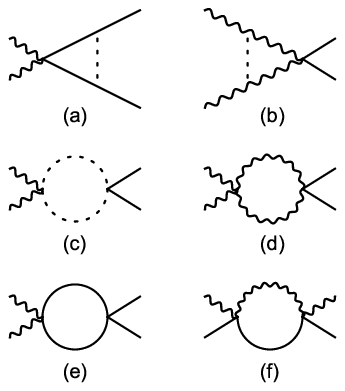}
\vspace{-0.3cm}
\caption{One-loop corrections to $\lambda_{\eta\phi}$.}\label{Fig_lambda_eta_phi}
\end{minipage}
\end{figure}

\begin{eqnarray}
\Delta S^{\alpha_h}_1
&=&\int^{b}\frac{d^{3}q}{(2\pi)^{3}}h_s(q)h_s(-q)\left\{\int^1_{b}\frac{d^{3}q'}
{(2\pi)^{3}}\left[-(3\gamma_h)^2G_h(q')G_h(q')-\gamma^2_{\eta h}G_\eta(q')G_\eta(q')
-\gamma^2_{\phi h}G_\phi(q')G_\phi(q')\right]\right\}\nonumber\\
&\approx&\int^{b}\frac{d^{3}q}{(2\pi)^{3}}h_s(q)h_s(-q)
\left[-\frac{(3\gamma_h)^2}{2\pi^2}\int^1_{b}d q'\left(\frac{1}{q'^2}
-\frac{4\alpha_h}{q'^4}\right)-\frac{\gamma^2_{\eta h}}{2\pi^2}\int^1_{b}
\frac{d q'}{q'^2}-\frac{\gamma^2_{\phi h}}{2\pi^2}\int^1_{b}d
q'\left(\frac{1}{q'^2}-\frac{4\alpha_\phi}{q'^4}\right)\right]\nonumber\\
&\approx&\int^{b}\frac{d^{3}q}{(2\pi)^{3}}h_s(q)h_s(-q)\left[-\frac{(3\gamma_h)^2}
{2\pi^2}\left(1-4\alpha_h\right)(-\ln b)-\frac{\gamma^2_{\eta h}}{2\pi^2}(-\ln b)
-\frac{\gamma^2_{\phi h}}{2\pi^2}\left(1-4\alpha_\phi\right)(-\ln b)\right]\nonumber\\
&=&-\int^{b}\frac{d^{3}q}{(2\pi)^{3}}h_s(q)h_s(-q)\frac{1}{2\pi^2}
\left[9\gamma^2_h(1-4\alpha_h)+\gamma^2_{\eta h}+\gamma^2_{\phi h}
(1-4\alpha_\phi)\right]l.\label{Eq_Delta_S_alpha_h_1}
\end{eqnarray}
Summing the diagrams in the second line of Fig.~\ref{Fig_alpha_h} gives rise to
\begin{eqnarray}
\Delta S^{\alpha_h}_2&=&\int^{b}\frac{d^{3}q}{(2\pi)^{3}}
h_s(q)h_s(-q)\frac{1}{2\pi^2}\left[3\beta_h(1-2\alpha_h)+\lambda_{h\phi}
(1-2\alpha_\phi)+\lambda_{\eta h}\right]l.\label{Eq_Delta_S_alpha_h_2}
\end{eqnarray}
\end{widetext}
The total one-loop correction to $\alpha_h$ is given by summing the
above contributions, namely
\begin{eqnarray}
\Delta S^{\alpha_h}
\!\!\!&=&\!\!\!\Delta S^{\alpha_h}_1+\Delta S^{\alpha_h}_2\nonumber\\
\!\!\!&=&\!\!\!\int^{b}\frac{d^{3}q}{(2\pi)^{3}}h_s(q)h_s(-q)\nonumber\\
&&\!\!\!\times\frac{1}{2\pi^2}\left[-9\gamma^2_h(1-4\alpha_h)-\gamma^2_{\eta h}
-\gamma^2_{\phi h}(1-4\alpha_\phi)\right.\nonumber\\
&&\!\!\!\left.+3\beta_h(1-2\alpha_h)+\lambda_{h\phi}(1-2\alpha_\phi)
+\lambda_{\eta h}\right]l.\label{Eq_Delta_S_alpha_h}
\end{eqnarray}
By combing the one-loop correction (\ref{Eq_Delta_S_alpha_h}) and
the free term proportional to $\alpha_h$, listed in Eq.
(\ref{Eq_action_eff_h}), we derive the RG equation of $\alpha_h$
after integrating fast modes and making scaling transformations,
\begin{eqnarray}
\frac{d\alpha_h}{dl} &=&
2\alpha_h-\frac{1}{4\pi^2}\Bigl[9\gamma^2_h(1-4\alpha_h)
+\gamma^2_{\phi h}(1-4\alpha_\phi)+\gamma^2_{\eta h} \nonumber\\
&& -\lambda_{\eta h}+3\beta_h(2\alpha_h-1) +
\lambda_{h\phi}(2\alpha_\phi-1)\Bigr],\label{RG_Eq_alpha_h}
\end{eqnarray}

By paralleling the above calculations, we can obtain the corrections
from other diagrams. Here, we only list the results by assigning
$\Delta^n_m\equiv \delta^{3}\left(\sum^n_{m=1} q_m\right)$,
\begin{eqnarray}
\Delta S^{\alpha_\phi}
\!\!\!&=&\!\!\!\int^{b}\frac{d^{3}q}{(2\pi)^{3}}\phi_s(q)\phi_s(-q)\nonumber\\
&&\!\!\!\times\frac{1}{2\pi^2}\Bigl\{-2\gamma^2_{\phi
h}[1-2(\alpha_\phi+\alpha_h)] +
(\lambda_{\eta\phi}+\lambda_{h\phi} \nonumber \\
&&\!\!\!+3\beta_\phi)-2(\alpha_h\lambda_{h\phi}+3\alpha_\phi
\beta_\phi)\Bigr\}l,\label{Eq_Delta_S_alpha_phi}\\
\Delta S^{\gamma_h}
\!\!\!&=&\!\!\!\int^{b}\prod^3_{m=1}\frac{d^3 q_m}{(2\pi)^{3}}h_s(q_m)\nonumber\\
&&\!\!\!\times\frac{\Delta^3_m}{2\pi^2}\Bigl[-36\gamma^3_{h}(1-6\alpha_h)
+9\beta_{h}\gamma_h(4\alpha_h-1)\nonumber\\
&&\!\!\!+\lambda_{h\phi}\gamma_{\phi h}
(4\alpha_\phi-1)-\lambda_{\eta h}\gamma_{\eta h}\Bigr]l,\label{Eq_Delta_S_gamma_h}\\
\Delta S^{\gamma_{\eta h}}
\!\!\!&=&\!\!\!\int^{b}\prod^3_{m=1}\frac{d^3 q_m}{(2\pi)^{3}}
\eta_s(q_1)\eta_s(q_2)h_s(q_3)\nonumber\\
&&\!\!\!\times\frac{\Delta^3_m}{2\pi^2}\Bigl\{-\gamma_{\eta h}
\Bigl[4\gamma^2_{\eta h}\left(1-2\alpha_h\right)
+3\beta_{\eta}\Bigr]+3\lambda_{\eta h}\nonumber\\
&&\!\!\!\times\gamma_h(4\alpha_h-1)
+\lambda_{\eta\phi}\gamma_{\phi h}(4\alpha_\phi-1)
\Bigr\}l,\label{Eq_Delta_S_gamma_eta_h}\\
\Delta S^{\gamma_{\phi h}}
\!\!\!&=&\!\!\!\int^{b}\prod^4_{m=1}\frac{d^3 q_m}{(2\pi)^{3}}
\phi_s(q_1)\phi_s(q_2)h_s(q_3)\nonumber\\
&&\!\!\!\times\frac{\Delta^4_m}{2\pi^2}\Bigl\{-4\gamma^3_{\phi h}
[1-2(2\alpha_\phi+\alpha_h)]\nonumber\\
&&\!\!\!+12(\alpha_h\lambda_{h\phi}\gamma_h
+\alpha_\phi\beta_{\phi}\gamma_{\phi h})\nonumber\\
&&\!\!\!-3(\lambda_{h\phi}\gamma_h+\beta_{\phi}\gamma_{\phi h})
-\lambda_{\eta\phi}\gamma_{\eta h}\Bigr\}l, \label{Eq_Delta_S_gamma_phi_h}\\
\Delta S^{\beta_h}
\!\!\!&=&\!\!\!\int^{b}\prod^3_{m=1}\frac{d^3 q_m}{(2\pi)^{3}}
h_s(q_1)h_s(q_2)h_s(q_3)h_s(q_4)\nonumber\\
&&\!\!\!\times\frac{\Delta^4_m}{2\pi^2}\Bigl[-108\beta_h
\gamma^2_h(1-6\alpha_h)-9\beta^2_h-\lambda^2_{\eta h} \nonumber\\
&&\!\!\!-\lambda^2_{h\phi}+4(9\alpha_h\beta^2_h+\alpha_\phi
\lambda^2_{h\phi})\Bigr]l, \label{Eq_Delta_S_beta_h}\\
\Delta S^{\beta_\eta} \!\!\! &=&
\!\!\!\int^{b}\prod^4_{m=1}\frac{d^3 q_m}{(2\pi)^{3}}
\eta_s(q_1)\eta_s(q_2)\eta_s(q_3)\eta_s(q_4)\nonumber\\
&&\!\!\!\times\frac{\Delta^4_m}{2\pi^2}\Bigl[-12\beta_\eta
\gamma^2_{\eta h}(1-2\alpha_h)-9\beta^2_\eta-\lambda^2_{\eta h}\nonumber\\
&&\!\!\!-\lambda^2_{\eta\phi}+4(\alpha_h\lambda^2_{\eta
h}+\alpha_\phi\lambda^2_{\eta\phi})\Bigr]l,\label{Eq_Delta_S_beta_eta}\\
\Delta S^{\beta_\phi} \!\!\!&=&\!\!\!\int^{b}\prod^4_{m=1}\frac{d^3
q_m}{(2\pi)^{3}}
\phi_s(q_1)\phi_s(q_2)\phi_s(q_3)\phi_s(q_4)\nonumber\\
&&\!\!\!\times\frac{\Delta^4_m}{2\pi^2}\Bigl\{-12\beta_\phi
\gamma^2_{\phi h}[1-2(2\alpha_\phi+\alpha_h)]-9\beta^2_\phi\nonumber\\
&&\!\!\!-\lambda^2_{h\phi}-\lambda^2_{\eta\phi}+4(9\alpha_\phi\beta^2_\phi
+\alpha_h\lambda^2_{h\phi})\Bigr\}l,\label{Eq_Delta_S_beta_phi}\\
\Delta S^{\lambda_{\eta h}}
\!\!\!&=&\!\!\!\int^{b}\prod^4_{m=1}\frac{d^3 q_m}{(2\pi)^{3}}
\eta_s(q_1)\eta_s(q_2)h_s(q_3)h_s(q_4)\nonumber\\
&&\!\!\!\times\frac{\Delta^4_m}{2\pi^2}\Bigl[-36\lambda_{\eta
h}\gamma^2_h(1-6\alpha_h)-4\lambda_{\eta h}\nonumber\\
&&\!\!\!\times\gamma^2_{\eta h}(1-2\alpha_h)+4(3\alpha_h
\lambda_{\eta h}\beta_h+\alpha_\phi\nonumber\\
&&\!\!\!\times\lambda_{\eta \phi}\lambda_{h \phi}
+4\alpha_h\lambda^2_{\eta h})-3\lambda_{\eta h}(\beta_h+\beta_\eta)\nonumber\\
&&\!\!\!-\lambda_{\eta \phi}\lambda_{h \phi}-8\lambda^2_{\eta h}\Bigr]l,
\label{Eq_Delta_S_lambda_eta_h}\\
\Delta S^{\lambda_{h\phi}}
\!\!\!&=&\!\!\!\int^{b}\prod^4_{m=1}\frac{d^3 q_m}{(2\pi)^{3}}
h_s(q_1)h_s(q_2)\phi_s(q_3)\phi_s(q_4)\nonumber\\
&&\!\!\!\times\frac{\Delta^4_m}{2\pi^2}\Bigl\{-4\lambda_{h\phi}
\gamma^2_{\phi h}[1-2(2\alpha_\phi+\alpha_h)]\nonumber\\
&&\!\!\!-36\lambda_{h\phi}\gamma^2_h(1-6\alpha_h)
+12\lambda_{h\phi}(\alpha_h\beta_h\nonumber\\
&&\!\!\!+\alpha_\phi\beta_\phi)+16\lambda^2_{h\phi}
(\alpha_h+\alpha_\phi)-3\lambda_{h\phi}\nonumber\\
&&\!\!\!\times(\beta_h+\beta_\phi)-\lambda_{\eta h}\lambda_{\eta\phi}
-8\lambda^2_{h\phi}\Bigr\}l, \label{Eq_Delta_S_lambda_h_phi}\\
\Delta S^{\lambda_{\eta\phi}}
\!\!\!&=&\!\!\!\int^{b}\prod^4_{m=1}\frac{d^3 q_m}{(2\pi)^{3}}
\eta_s(q_1)\eta_s(q_2)\phi_s(q_3)\phi_s(q_4)\nonumber\\
&&\!\!\!\times\frac{\Delta^4_m}{2\pi^2}\Bigl\{-4\lambda_{\eta\phi}
\gamma^2_{\phi h}[1-2(2\alpha_\phi+\alpha_h)]\nonumber\\
&&\!\!\!-4\lambda_{\eta\phi}\gamma^2_{\eta h}(1-2\alpha_h)+4(\alpha_h
\lambda_{\eta h}\lambda_{h\phi}\nonumber\\
&&\!\!\!+3\alpha_\phi\lambda_{\eta\phi}\beta_\phi+4\alpha_\phi
\lambda^2_{\eta\phi})-3\lambda_{\eta\phi}(\beta_\eta+\beta_\phi)\nonumber\\
&&\!\!\!-\lambda_{\eta h}\lambda_{h\phi}
-8\lambda^2_{\eta\phi}\Bigr\}l.\label{Eq_Delta_S_lambda_eta_phi}
\end{eqnarray}
Eqs.~(\ref{Eq_Delta_S_alpha_h})~-~(\ref{Eq_Delta_S_lambda_eta_phi})
represent one-loop corrections to all the mass and interaction terms
of the effective action. Now one can add these corrections to the
original action terms, and obtain an effective new action.

\subsection{RG equations}

Based on the above calculations, it is now straightforward to write
down the coupled RG equations for all the
parameters~\cite{Wang2013NJP}. Performing calculations that lead to
Eq. (\ref{RG_Eq_alpha_h}), we have
\begin{widetext}
\begin{eqnarray}
\frac{d\alpha_h}{dl}
&=&2\alpha_h-\frac{1}{4\pi^2}\Bigl[9\gamma^2_h(1-4\alpha_h)
+\gamma^2_{\phi h}(1-4\alpha_\phi)+\gamma^2_{\eta h} -\lambda_{\eta h}
+3\beta_h(2\alpha_h-1)+\lambda_{h\phi}(2\alpha_\phi-1)\Bigr],\nonumber\\
\frac{d\alpha_\phi}{dl}
&=&2\alpha_\phi-\frac{1}{4\pi^2}\Bigl\{2\gamma^2_{\phi h}
\Bigl[1-2(\alpha_\phi+\alpha_h)-\lambda_{\eta\phi}\Bigr]
+\lambda_{h\phi}(2\alpha_h-1)-3\beta_\phi(1-2\alpha_\phi)\Bigr\},\nonumber\\
\frac{d\gamma_h}{dl}
&=&\frac{3}{2}\gamma_h+\frac{1}{2\pi^2}
\Bigl\{9\gamma_h\Bigl[\beta_{h}(4\alpha_h-1)
-4\gamma^2_{h}(1-6\alpha_h)\Bigr]+\lambda_{h\phi}
\gamma_{\phi h}(4\alpha_\phi-1)-\lambda_{\eta h}
\gamma_{\eta h}\Bigr\},\nonumber\\
\frac{d\gamma_{\eta h}}{dl}
&=&\frac{3}{2}\gamma_{\eta h}+\frac{1}{2\pi^2}
\Bigl\{3\lambda_{\eta h}\gamma_h(4\alpha_h-1)
+\lambda_{\eta\phi}\gamma_{\phi h}(4\alpha_\phi-1)
-\gamma_{\eta h}\Bigl[3\beta_{\eta}+4\gamma^2_{\eta h}
(1-2\alpha_h)\Bigr]\Bigr\},\nonumber\\
\frac{d\gamma_{\phi h}}{dl}
&=&\frac{3}{2}\gamma_{\phi h}+\frac{1}{2\pi^2}
\Bigl\{3\lambda_{h\phi}\gamma_h(4\alpha_h
-1)+3\beta_{\phi}\gamma_{\phi h}(4\alpha_\phi-1)
-\lambda_{\eta\phi}\gamma_{\eta h}-4\gamma^3_{\phi h}
\Bigl[1-2(2\alpha_\phi+\alpha_h)\Bigr]\Bigr\},\nonumber\\
\frac{d\beta_h}{dl}
&=&\beta_h+\frac{1}{\pi^2}\Bigl\{9\beta_h
\Bigl[\beta_h(4\alpha_h-1)-12\gamma^2_h
(1-6\alpha_h)\Bigr]+\lambda^2_{h\phi}
(4\alpha_\phi-1)-\lambda^2_{\eta h}\Bigr\},\nonumber\\
\frac{d\beta_\eta}{dl}
&=&\beta_\eta+\frac{1}{\pi^2}\Bigl\{\lambda^2_{\eta\phi}
(4\alpha_\phi-1)+\lambda^2_{\eta h}(4\alpha_h-1)
-3\beta_\eta\Bigl[3\beta_\eta+4\gamma^2_{\eta h}
(1-2\alpha_h)\Bigr]\Bigr\},\label{Eq_effective_RG}\\
\frac{d\beta_\phi}{dl}
&=&\beta_\phi+\frac{1}{\pi^2}\Bigl\{3\beta_\phi
\Bigl[12\alpha_\phi\beta_\phi-3\beta_\phi
-4\gamma^2_{\phi h}\Bigl(1-2(2\alpha_\phi
+\alpha_h)\Bigr)\Bigr]+\lambda^2_{h\phi}(4\alpha_h-1)
-\lambda^2_{\eta\phi}\Bigr\},\nonumber\\
\frac{d\lambda_{\eta h}}{dl}
&=&\lambda_{\eta
h}+\frac{1}{2\pi^2}\Bigl\{\lambda_{\eta h}
\Bigl[4\alpha_h(3\beta_h+4\lambda_{\eta h})-8\lambda_{\eta
h}-3(\beta_h+\beta_\eta)-36\gamma^2_h(1-6\alpha_h)
-4\gamma^2_{\eta h}\nonumber\\
&&\times(1-2\alpha_h)\Bigr]+\lambda_{\eta \phi}
\lambda_{h \phi}(4\alpha_\phi-1)\Bigr\},\nonumber\\
\frac{d\lambda_{h \phi}}{dl}
&=&\lambda_{h\phi}+\frac{1}{2\pi^2}
\Bigl\{\lambda_{h\phi}\Bigl[12(\alpha_h\beta_h+\alpha_\phi\beta_\phi)
-3(\beta_h+\beta_\phi)-4\gamma^2_{\phi h}\Big(1-2(2\alpha_\phi
+\alpha_h)\Bigr)\nonumber\\
&&-36\gamma^2_h(1-6\alpha_h)+8\lambda_{h\phi}
\Big(2(\alpha_h+\alpha_\phi)-1\Bigr)\Bigr]
-\lambda_{\eta h}\lambda_{\eta\phi}\Bigr\},\nonumber\\
\frac{d\lambda_{\eta \phi}}{dl}
&=&\lambda_{\eta \phi}+\frac{1}{2\pi^2}\Bigl\{\lambda_{\eta\phi}
\Bigl[4\alpha_\phi(3\beta_\phi+4\lambda_{\eta\phi})
-3(\beta_\eta+\beta_\phi)-4\gamma^2_{\eta h}(1-2\alpha_h)\nonumber\\
&&-4\gamma^2_{\phi h}\Bigl(1-2(2\alpha_\phi+\alpha_h)\Bigr)
-8\lambda_{\eta\phi}\Bigr]+\lambda_{\eta h}\lambda_{h\phi}
(4\alpha_h-1)\Bigr\}.\nonumber
\end{eqnarray}

As mentioned in Sec.~\ref{Sec_eff_theory}, there are indeed only
five fundamental parameters: $\alpha$, $r$, $\beta$, $u$, and
$\lambda$. To obtain the running behavior of the system, we can
either directly solve Eqs.~(\ref{Eq_effective_RG}), or solve the RG
equations of fundamental parameters extracted from
Eqs.~(\ref{Eq_effective_RG}). We have verified that these two
methods lead to the same conclusion. Here we choose to adopt the
second method mainly for two reasons. First, the low-energy behavior
of the original action Eqs.(1-4) can be most clearly seen from the
$l$-dependence of the five fundamental parameters. In addition, it
is technically easier to display the detailed $l$-dependence of five
parameters than eleven parameters. The RG equations for $\alpha$,
$r$, $\beta$, $u$, and $\lambda$ extracted from
Eq.~(\ref{Eq_effective_RG}) are
\begin{eqnarray}
\frac{d\alpha}{dl}
&=&2\alpha-\frac{1}{4\pi^2}\left\{\alpha\beta
(5-18\alpha)+\frac{\beta}{2}(3\alpha-2)+\frac{2\alpha\lambda^2}{\beta}
\left[1-4\left(\frac{\lambda\alpha}{\beta}+r\right)\right]
+\frac{\lambda}{2}\left[2\left(\frac{\lambda\alpha}
{\beta}+r\right)-1\right]\right\},\nonumber\\
\frac{dr}{dl}
&=&2\left(\frac{\lambda\alpha}{\beta} +
r\right) - \frac{1}{4\pi^2}\left\{\frac{2\lambda\alpha}{\beta}
\left[2(1-2\alpha) - 4\left(\frac{\lambda\alpha}{\beta} +
r\right) - \lambda\right]-3u\left[1-2\left(\frac{\lambda\alpha}{\beta}
+r\right)\right]\right.\nonumber \\
&&\left.+\frac{\lambda}{2}(2\alpha-1)\right\}+\left(\frac{\lambda\alpha}
{\beta^2}\frac{d\beta}{dl}-\frac{\lambda}{\beta}
\frac{d\alpha}{dl}-\frac{\alpha}{\beta}\frac{d\lambda}{dl}\right),\nonumber\\
\frac{d\beta}{dl}
&=&\beta + \frac{1}{\pi^2}\left\{9\beta^2\left[\frac{(4\alpha-1)}{4}
-6\alpha(1-6\alpha)\right] + \lambda^2\left[4\left(\frac{\lambda\alpha}
{\beta}+r\right)-1\right]-\frac{\beta^2}{4}\right\},\label{Eq_Renormalized_RG}\\
\frac{du}{dl}
&=&u + \frac{1}{\pi^2}\left\{3u\left[\left(12u+\frac{32\alpha\lambda^2}
{\beta}\right)\left(\frac{\lambda\alpha}{\beta}+r\right)-3u
-\frac{8\alpha\lambda^2}{\beta}(1-2\alpha)\right]
+\frac{\lambda^2}{2}(2\alpha-1)\right\},\nonumber\\
\frac{d\lambda}{dl}
&=&\lambda+\frac{1}{\pi^2}\left\{\frac{\lambda}{2}\left[4\left(3u
+\frac{8\lambda^2\alpha}{\beta} +2\lambda\right)\left(\frac{\lambda
\alpha}{\beta}+r\right)-3\left(\frac{\beta}{4}+u\right)-
3\alpha\beta(5-36\alpha)\right.\right.\nonumber \\
&&\left.\left.+4\left(\lambda + \frac{2\lambda^2\alpha}
{\beta}\right)(2\alpha-1)\right]-\frac{\beta\lambda}
{8}\right\}.\nonumber
\end{eqnarray}
These flow equations are strongly coupled to each other. We know
from RG theory that only stable fixed points can be realized in the
thermodynamic limit. So the next step is to find the possible fixed
point, which by definition should be unchanged under RG
transformations. In the next section, we will solve these equations
self-consistently.

\begin{figure}
   \centering
   \hspace{-1.0cm}
   \epsfig{file=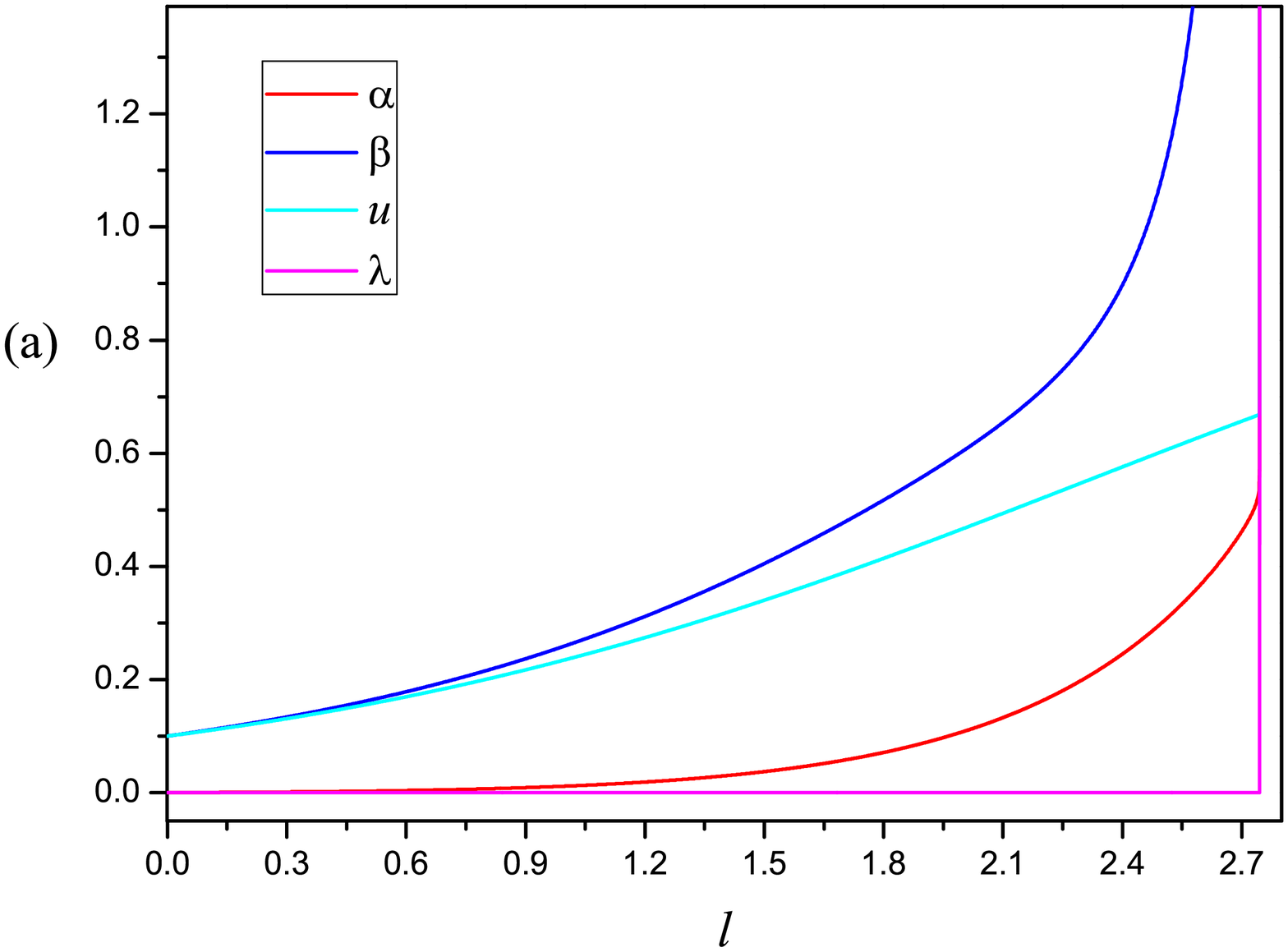,height = 6.15cm,width=9cm}\hspace{-1.70cm}
   \epsfig{file=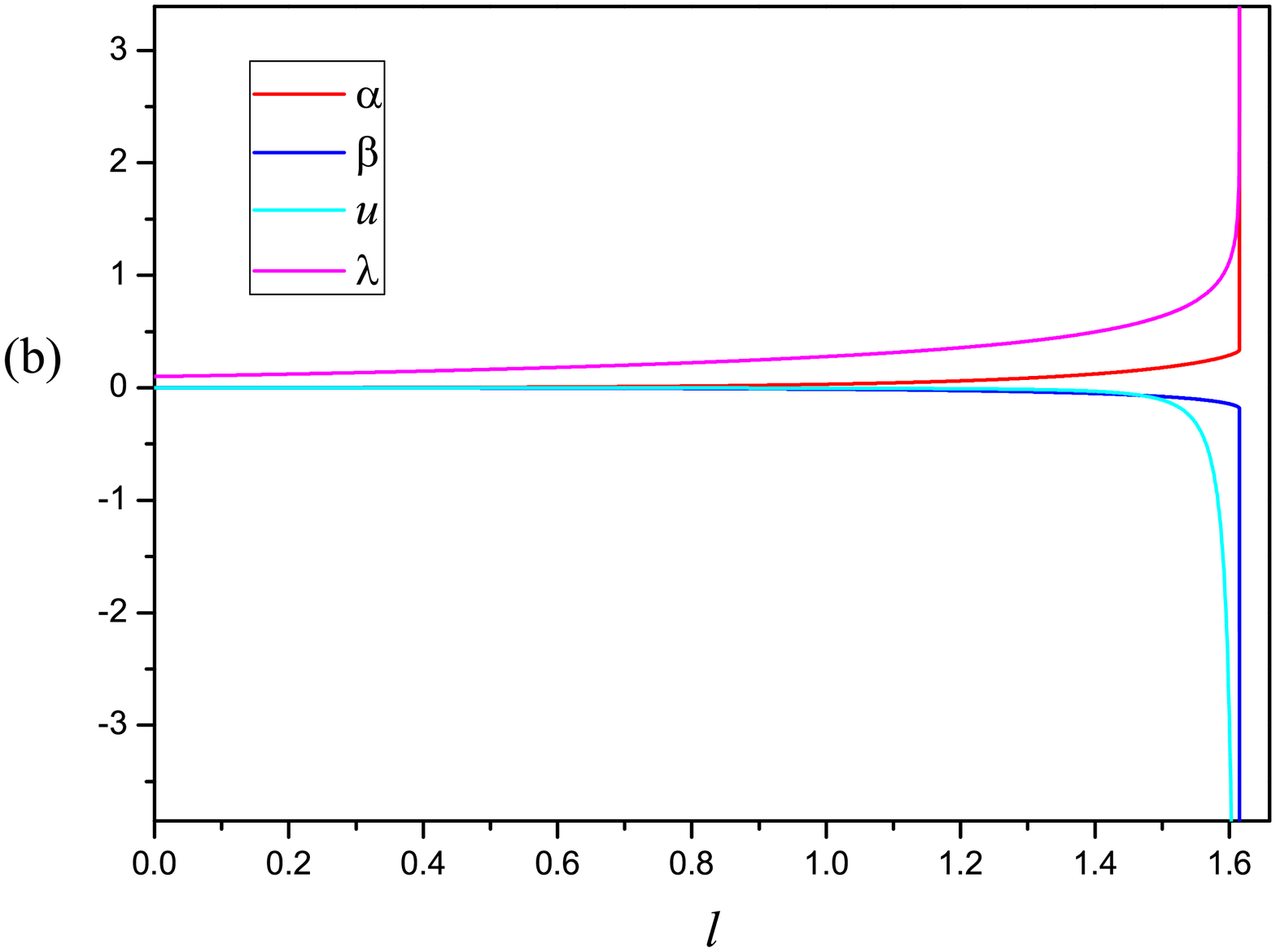,height = 6.15cm,width=9cm}\\ \vspace{-0.80cm}
   \hspace{-1.0cm}
   \epsfig{file=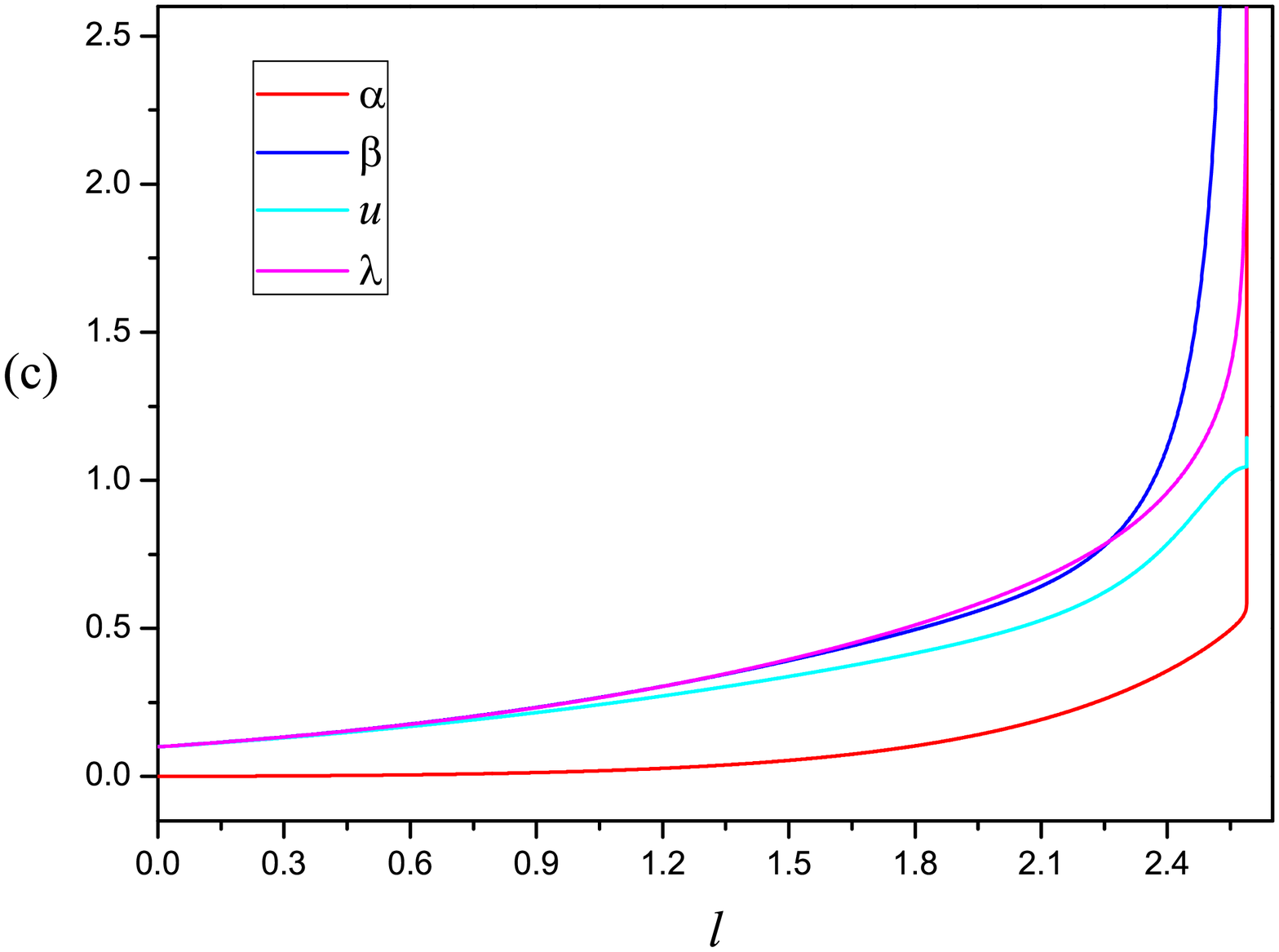,height = 6.15cm,width=9cm}\hspace{-1.70cm}
   \epsfig{file=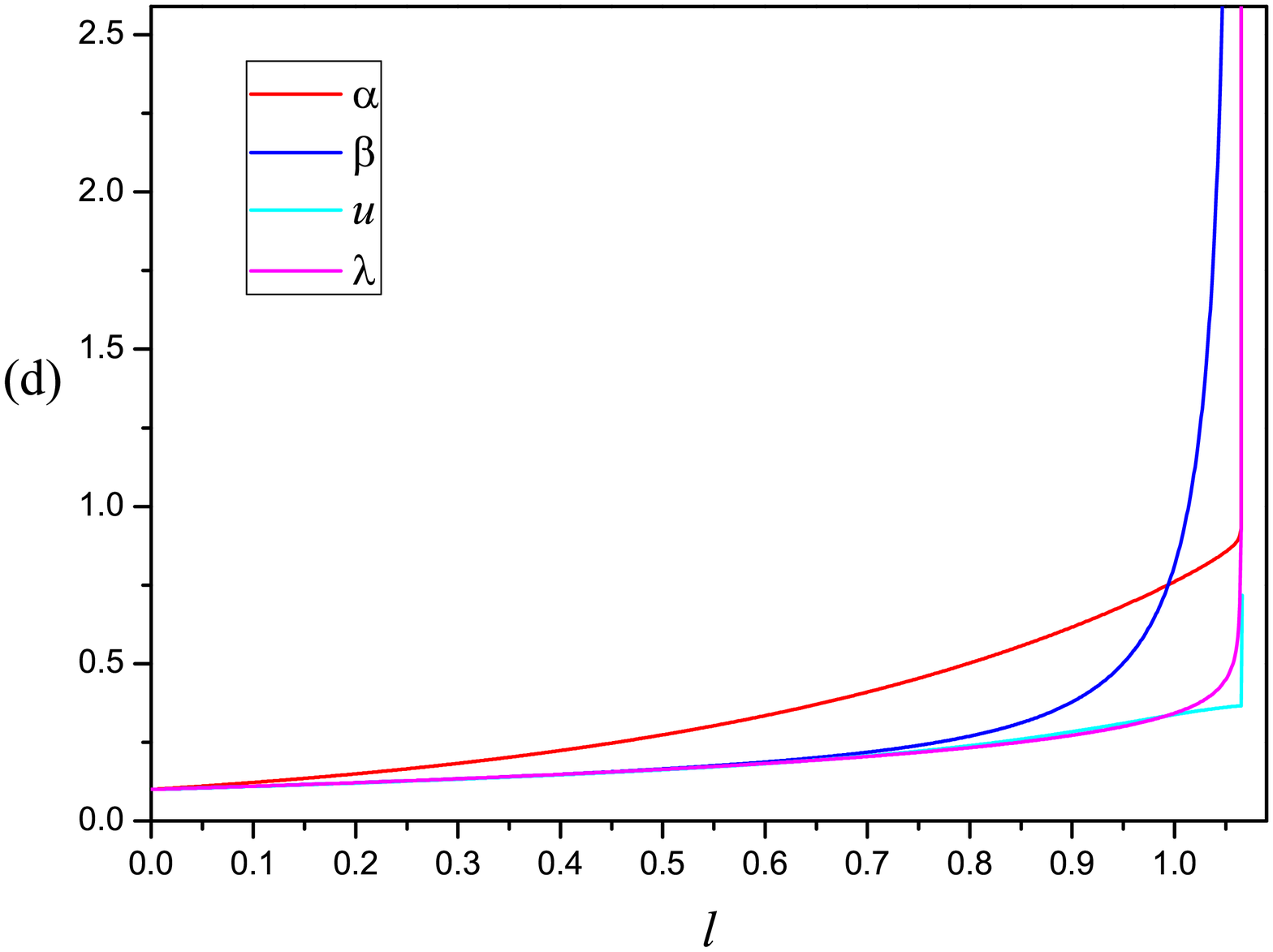,height = 6.15cm,width=9cm}\\
\vspace{-0.35cm} \caption{Initial values of dimensionless
parameters: (a) $\alpha_0=10^{-7}$, $\beta_0=u_0=10^{-1}$,
$\lambda_0=10^{-7}$; (b) $\alpha_0=10^{-7}$, $\beta_0=u_0=10^{-7}$,
$\lambda_0=10^{-1}$; (c) $\alpha_0=10^{-7}$, $\beta_0=u_0=10^{-1}$,
$\lambda_0=10^{-1}$; (d) $\alpha_0=10^{-1}$, $\beta_0=u_0=10^{-1}$,
$\lambda_0=10^{-1}$.}\label{Fig_Running_away}
\end{figure}

\begin{figure}
   \centering
   \hspace{-1.0cm}
   \epsfig{file=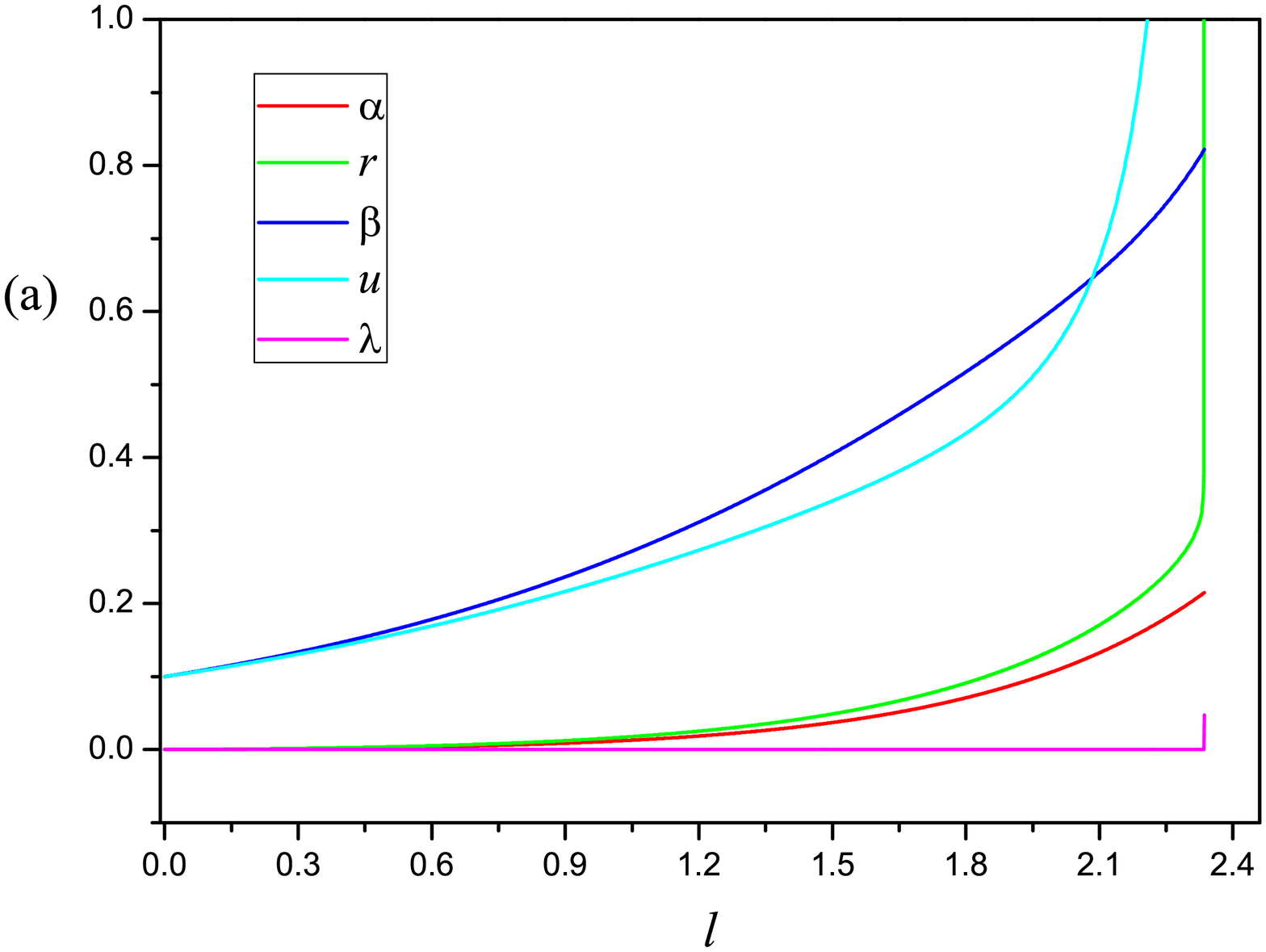,height = 6.15cm,width=9cm}\hspace{-1.70cm}
   \epsfig{file=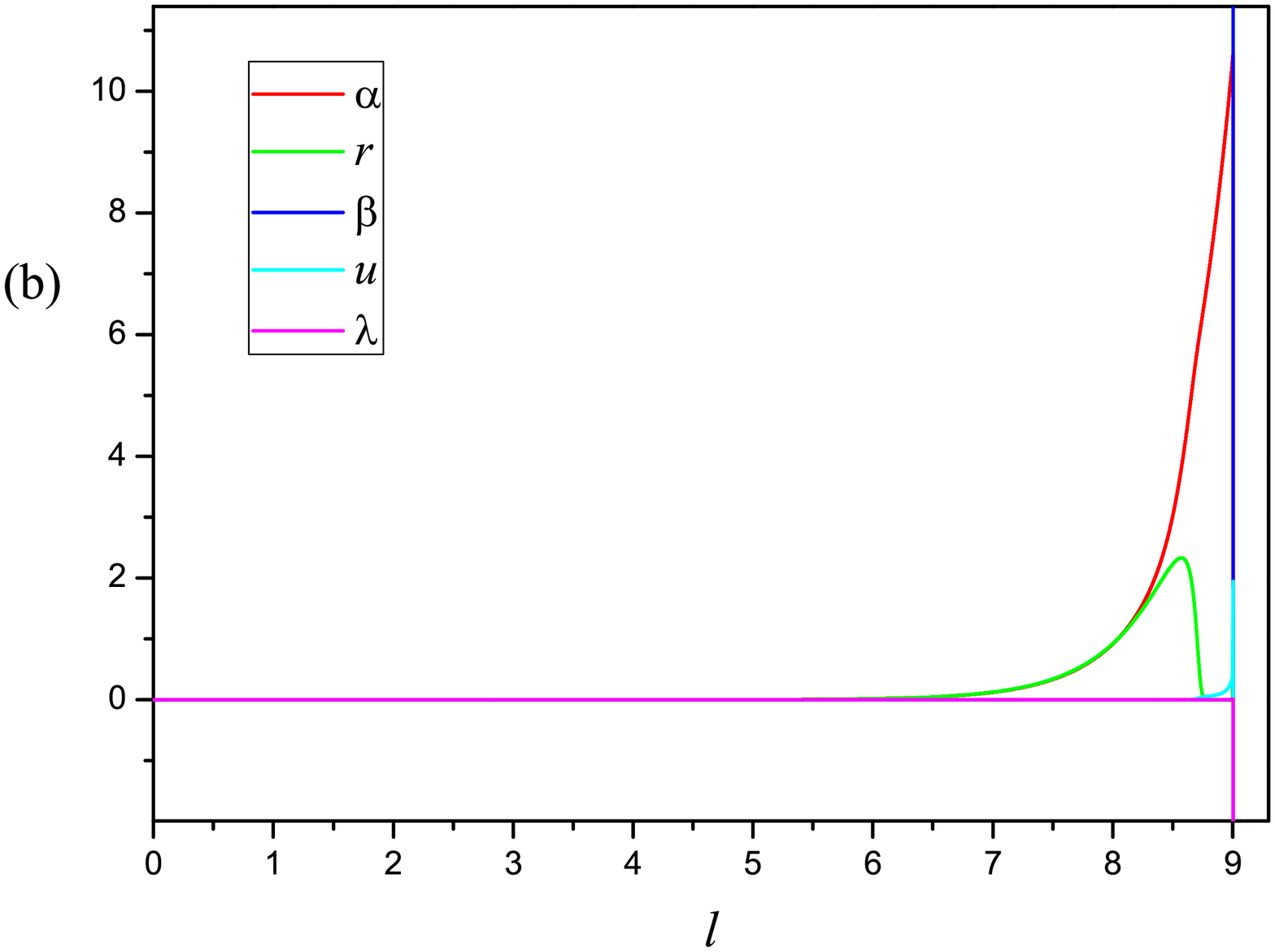,height = 6.15cm,width=9cm}\\ \vspace{-0.80cm}
   \hspace{-1.0cm}
   \epsfig{file=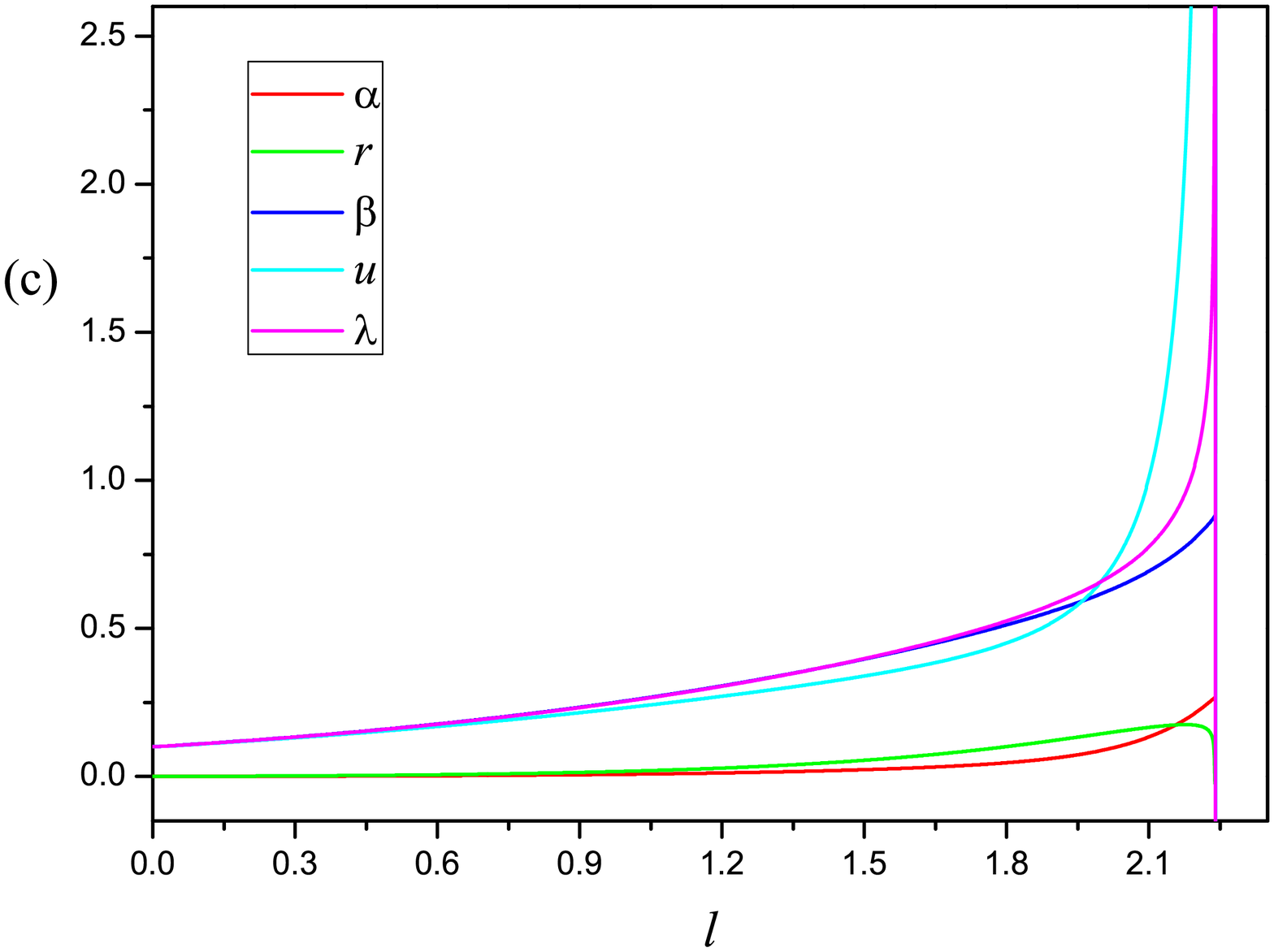,height = 6.15cm,width=9cm}\hspace{-1.70cm}
   \epsfig{file=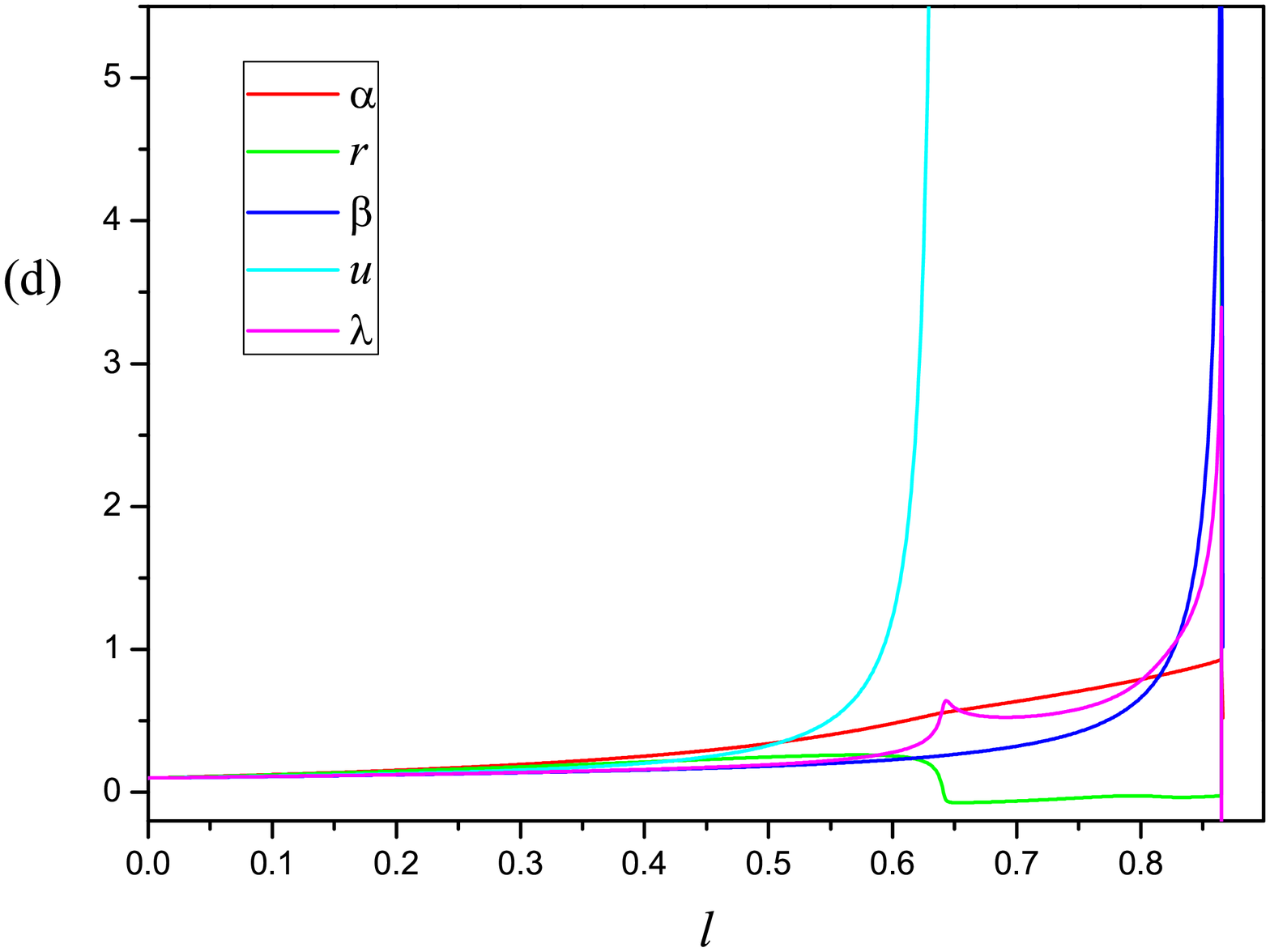,height = 6.15cm,width=9cm}\\
\vspace{-0.35cm}
\caption{Initial values of dimensionless parameters:
(a) $\alpha_0=r_0=10^{-7}$, $\beta_0=u_0=10^{-1}$,$\lambda_0=10^{-7}$; (b) $\alpha_0=r_0=10^{-7}$,
$\beta_0=u_0=10^{-7}$, $\lambda_0=10^{-7}$; (c) $\alpha_0=r_0=10^{-7}$, $\beta_0=u_0=10^{-1}$,
$\lambda_0=10^{-1}$; (d) $\alpha_0=r_0=10^{-1}$, $\beta_0=u_0=10^{-1}$,
$\lambda_0=10^{-1}$.}\label{Fig_Running_away_2}
\end{figure}
\end{widetext}

\section{RG Solutions and analysis}\label{Sec_RG_Solutions}

In this section, we study the RG equations obtained in the last
section to examine whether there is any stable nontrivial fixed
point in the presence of finite interactions. For this purpose, we
will go through the following two steps. First, we require all the
RG flow equations to vanish, leading to a set of coupled
differential equations. Second, we solve these equations numerically
and judge whether the solutions are stable as the running scale $l$
goes to infinity.

At the quantum critical point $x_2$, $r$ vanishes. By taking all the
RG equations (\ref{Eq_Renormalized_RG}) to vanish, one can check
that there is no physical fixed point. To confirm this result, we
next solve Eqs.~(\ref{Eq_Renormalized_RG}) numerically at $r=0$ and
extract the explicit $l$-dependence of the other four parameters.
The stability of the system is mainly determined by the behaviors of
these parameters in the limit $l \rightarrow \infty$. After choosing
certain initial (bare) values for $\alpha$, $\beta$, $u$, and
$\lambda$, we obtain the $l$-dependence of renormalized parameters
and show the results in Fig.~\ref{Fig_Running_away}. It can be
seen from Fig.~\ref{Fig_Running_away} that the qualitative
conclusion does not change as the initial values of the parameters
vary. In particular, the coupling parameters $\beta$, $u$, and
$\lambda$ diverge rapidly as $l$ grows. These runaway behaviors
clearly show the absence of any stable fixed point, and strongly
suggest that the system undergoes first-order transitions~
\cite{Domany_Mukamel_Fishe, Chen_Lubensky_Nelson, Rudnick1978PRB,
Iacobson_Amit, Cardy1996, She} as a consequence of ordering competition.
Moreover, we find that this conclusion holds even at $r\neq 0$. In the
case $r < 0$, $\phi$ is also in the ordered phase, and one needs to
expand $\phi$ in a form similar to Eq.~(\ref{Eq_psi_renormalized}) if the
corresponding broken symmetry is also continuous. The effective field
theory would become much more complicated, but the analysis can be
performed in exactly the same way.

We next compare our results with previous work. The competition
between two distinct order parameters was investigated within an
effective (3+1)-dimensional field theory in Refs.~\cite{Kosterlitz,
She}. Detailed RG calculations revealed a stable fixed point, called
biconical fixed point, in the system that contains a two-component
order parameter and a one-component order parameter~\cite{Kosterlitz}.
Our work differs from previous one mainly in two aspects. First, the
complex order parameter is assumed to stay in the ordered phase in
our work, whereas the real order parameter is very close to its
quantum critical point, which allows us to carefully examine the
impacts of quantum critical fluctuation of order parameter. Second,
the effect of massless Goldstone boson generated by continuous
symmetry breaking is explicitly incorporated.

We now wish to figure out the factor that drives the first-order
transition. In particular, is the runaway behavior triggered by the
massless Goldstone boson? To clarify this point, we still separate
the mean value and fluctuation of order parameter in the ordered
state, but choose order parameter $\psi$ to be a real scalar field
which is formed by breaking a discrete symmetry. Therefore, the
current system does not contain Goldstone bosons. After analogous
calculations, we did not find any stable fixed point, so the phase
transition is still first-order. This does not mean that Goldstone
boson is unimportant. As a massless excitation (particle), Goldstone
boson should have important influence on the physical properties of
the system. In the present problem, however, it turns out that the
amplitude fluctuation alone is significant enough to lead to runaway
behavior.

\section{Coexisting region of two competing orders}\label{Sec_Coexisting_region}

In the previous sections, we have considered the disordered phase
and the quantum critical point of long-range order $\phi$. For
completeness, we now turn to the coexistence region where both order
parameters $\psi$ and $\phi$ have finite mean values. In this
region, $\phi$ should also be decomposed as
\begin{eqnarray}
\phi = c + \sqrt{\frac{r}{u}},\label{Eq_phi_renormalized}
\end{eqnarray}
with $\langle c\rangle=0$. Now the effective action becomes
\begin{eqnarray}
\mathcal{L} &=& \frac{1}{2}(\partial_\mu h)^2
+\alpha_hh^2+\gamma_hh^3 + \frac{\beta_h}{2}h^4+
\frac{1}{2}(\partial_\mu\eta)^2\nonumber\\
&&+\alpha_\eta\eta^2+\frac{\beta_\eta}{2}\eta^4+
\frac{1}{2}(\partial_\mu c)^2 + \alpha_c c^2 + \gamma_c c^3 +
\frac{\beta_c}{2}c^4\nonumber\\
&& +\gamma_{hc}h c + \gamma_{h^2c}h^2 c + \gamma_{\eta^2c}\eta^2 c +
\gamma_{c^2h}c^2 h + \gamma_{\eta^2 h}\eta^2 h\nonumber\\
&&+ \lambda_{hc}h^2 c^2 + \lambda_{\eta c}\eta^2 c^2
+ \lambda_{h\eta}h^2 \eta^2,\label{Eq_L_total_2}
\end{eqnarray}
where
\begin{eqnarray}
\alpha_h&=&\alpha+\frac{\lambda r}{2u},\,\,\alpha_\eta=\frac{\lambda r}{2u},
\,\,\alpha_c = 2r+\frac{\lambda\alpha}{\beta},\\
\beta_c&=&u,\,\,\beta_h=\beta_\eta=\frac{\beta}{4},\,\,\gamma_h=\gamma_{\eta^2h}
=\frac{\sqrt{2\alpha\beta}}{2},\\
\gamma_c&=&2\sqrt{u r},\,\,\gamma_{hc}=2\lambda\sqrt{\frac{2\alpha r}{\beta u}},\gamma_{c^2h}=\lambda\sqrt{\frac{2\alpha}{\beta}},\\
\gamma_{h^2c}&=&\gamma_{\eta^2c}=\lambda\sqrt{\frac{r}{u}},\lambda_{h\eta}=\frac{\beta}{4},
\lambda_{hc}=\lambda_{\eta c}=\frac{\lambda}{2}.
\end{eqnarray}
An interesting new result is that the originally massless Goldstone
boson $\eta$ acquires a finite mass due to the competitive
interaction between order parameters $\psi$ and $\phi$. This
mass-generating mechanism is a spectacular feature of ordering
competition. Apparently, it is physically very different from
Anderson-Higgs mechanism, because the latter relies crucially on the
presence of spontaneous local gauge symmetry breaking which in
general does not exit in our case (as long as $\psi$ does not
correspond to a superconducting order parameter). The effective
Goldstone boson mass vanishes at the quantum critical point $x_2$
since $\alpha_{\eta}\propto \lambda r \rightarrow 0$ as
$r\rightarrow 0$. Furthermore, $\alpha_{\eta}$ vanishes when the
competing orders decouple from each other, i.e.,
$\alpha_{\eta}\propto \lambda r \rightarrow 0$ as
$\lambda\rightarrow 0$

Now there are no massless modes in the effective action
(\ref{Eq_L_total_2}). Such an action can be analyzed in exactly the
same way as that presented in Sec.~(\ref{Sec_RG_Calculations}).
After tedious but straightforward calculations, we obtain the RG
equations of five fundamental parameters, $\alpha$, $r$, $\beta$,
$u$, and $\gamma$, which will not be explicitly shown here due to
the formal complicity. By carrying out numerical calculations, we
show the running of these parameters in
Fig.~\ref{Fig_Running_away_2}. It is clear that the phase
transitions become first-order even in the coexisting region of
competing orders.

\section{Summary}\label{Sec_Summary}

In summary, we have carried out a RG analysis within a
(2+1)-dimensional quantum field theory composed of two scalar
fields, which is able to describe the interplay between two distinct
order parameters. Different from previous treatments, we separate
the quantum fluctuations of amplitude and phase of complex order
parameter in the ordered state, and study their impacts on the
stability of the system respectively. After deriving and analyzing
the RG flow equations of all the relevant parameters, we have
demonstrated that the phase transitions become first-order due to
the absence of stable fixed point. We also have shown that this
conclusion holds in both the ordered and disordered phases, and also
at the quantum critical point.

The RG calculations presented in this paper are valid only for weak
couplings. If the competitive interaction between distinct orders is
not weak, one needs to invoke strong coupling approach. For
instance, we might assume the scalar field has a large flavor $N$
and then perform $1/N$-expansion. In addition, when applied to study
quantum phase transition, the scalar field may have a nontrivial
dynamical exponent $z\neq1$ so that its propagator is of the form
$\frac{1}{k^2 + k_0/k^{z-2}}$ \cite{She}. In this case, the RG
transformations would be different from the case with $z = 1$. It is
also interesting to include the interaction between scalar field and
fermionic degrees of freedom, which are known to important in the
theoretical description of competing orders \cite{Wang2013NJP}.

\vspace{0.5cm}
\section{Acknowledgments}

We would like to thank Jing-Rong Wang for helpful discussions. J.W.
is supported by China Postdoctoral Science Foundation under Grant
No.2014M560510. G.Z.L. is supported by the National Natural Science
Foundation of China under Grant No.11274286.


\end{document}